\documentclass[lineno]{JFM-FLM_Au}

\definecolor{ReBlue}{RGB}{0,112,192}
\definecolor{ReRed}{RGB}{192,0,0}

\newcommand{\plotstyle}[2]{%
  \begingroup
  \color{#1}%
  \raisebox{-0.15ex}{#2}%
  \hspace{0.45em}%
  \rule[0.5ex]{0.8cm}{0.9pt}%
  \endgroup
}

\usepackage[dvipsnames]{xcolor}

\usepackage{enumitem}

\lefttitle{J. Hwang}
\righttitle{Journal of Fluid Mechanics}
\title{Single-eddy contributions to streamwise velocity variance in turbulent pipe flow. Part 1. The single-eddy intensity function.}
\author{Jinyul Hwang}
\affiliation{School of Mechanical Engineering, Pusan National University, 2 Busandaehak-ro 63beon-gil, Geumjeong-gu, Busan 46241, Republic of Korea
}
\corresau{Jinyul Hwang, \email{jhwang@pusan.ac.kr}}

\begin{document}
\nolinenumbers
\maketitle

\begin{abstract}
Townsend’s attached-eddy hypothesis explains the asymptotic statistics of wall turbulence through a superposition of self-similar attached eddies.
At a given observation location $y$, their active portions contribute to turbulent momentum transfer, whereas their inactive portions retain only wall-parallel velocity fluctuations.
Since this distinction depends on $y$ relative to the eddy height $y_l$, resolving it requires fixing $y_l$ and tracing the eddy contribution across $y^*=y/y_l$.
Single-eddy intensity functions describe such contributions but are typically prescribed a priori.
Here, we extract these functions from DNS data of turbulent pipe flow at $Re_\tau\simeq930$--$6000$ and examine how active and inactive portions of wall-coherent motions contribute to the inner-scaled streamwise turbulence intensity $\langle uu \rangle^+$.
Differencing cumulative wall-coherent spectra obtained using spectral linear stochastic estimation isolates contributions at prescribed $y_l$, whose wavenumber integration yields the streamwise and wall-normal intensity functions ($I_{uu}$ and $I_{vv}$).
The $I_{uu}$ and $I_{vv}$ profiles each collapse across $y_l$ and share a peak at $y_a^*$, close to the von K\'arm\'an constant.
Near $y_a^*$, both components remain significant, indicating the active portion. 
By contrast, for $y^*\ll y_a^*$, $I_{vv}$ diminishes while $I_{uu}$ remains finite, indicating the inactive portion.
Over the self-similar range, both peak intensities scale as $y_l^{-1}$, and at fixed $y^*$, $I_{uu}$ retains this scaling in the active and inactive portions over their respective $y_l$ ranges.
Active-portion integration yields an approximately constant contribution, consistent with the universality of active motions, whereas integration of the inactive portion recovers the classical near-wall logarithmic variation.
An outer logarithmic tendency with a slope close to the Townsend--Perry constant can arise from integrating the active contributions up to their outer cutoff.
At the near-wall peak of $\langle uu\rangle^+$, the scaling steepens to $I_{uu} \sim y_l^{-(1+\beta)}$, with $\beta\simeq1/4$, yielding a finite asymptote with a $Re_\tau^{-1/4}$ defect. 
The additional exponent is attributed to the viscous attenuation within inactive footprints, associated with a $y_l$-dependent dissipative scale.
The wall-coherent motions also produce a broad outer shoulder, while a localized residual related to wall-incoherent motions may contribute to a secondary peak. 
A possible geometrical interpretation of $y_a^*$ and the relation of the logarithmic variations to the Townsend--Perry constant are also discussed.

\end{abstract}

\begin{keywords}
turbulent flows, turbulent boundary layers, turbulence simulation
\end{keywords}

\section{Introduction}
\label{sec:Introduction}

The attached-eddy hypothesis (AEH) proposed by \citet{Townsend61,Townsend76} provides a framework for understanding the asymptotic statistical behaviour of wall turbulence in terms of energy-containing motions.
At sufficiently high Reynolds numbers, an overlap region emerges that is separated from both the viscous near-wall and outer regions.
In the overlap region, the Reynolds shear stress is approximately constant ($\approx u_\tau^2$, where $u_\tau$ is the friction velocity), while turbulence production and dissipation are roughly in balance.
Thus, the region is referred to as the constant-stress equilibrium layer \citep{Townsend61}.
Townsend suggested that the energy-containing motions in this region are characterized by $u_\tau$ and the wall-normal distance $y$, independently of viscosity.
Since their characteristic size scales with $y$, these motions are called attached eddies (AEs).
The second-order moments associated with an individual AE of height $y_l$ are described by the single-eddy intensity functions (i.e., its contributions to the Reynold stresses).
A linear superposition of these functions over eddy heights, with a statistical weighting proportional to $y_l^{-1}$, predicts logarithmic variations of the streamwise and spanwise turbulence intensities and a constant wall-normal turbulence intensity in the constant-stress layer.
In the present work, we focus on these single-eddy intensity functions.

\begin{figure}
\centerline{\includegraphics[trim=5.3cm 2.8cm 5.3cm 2.8cm,width=11cm]{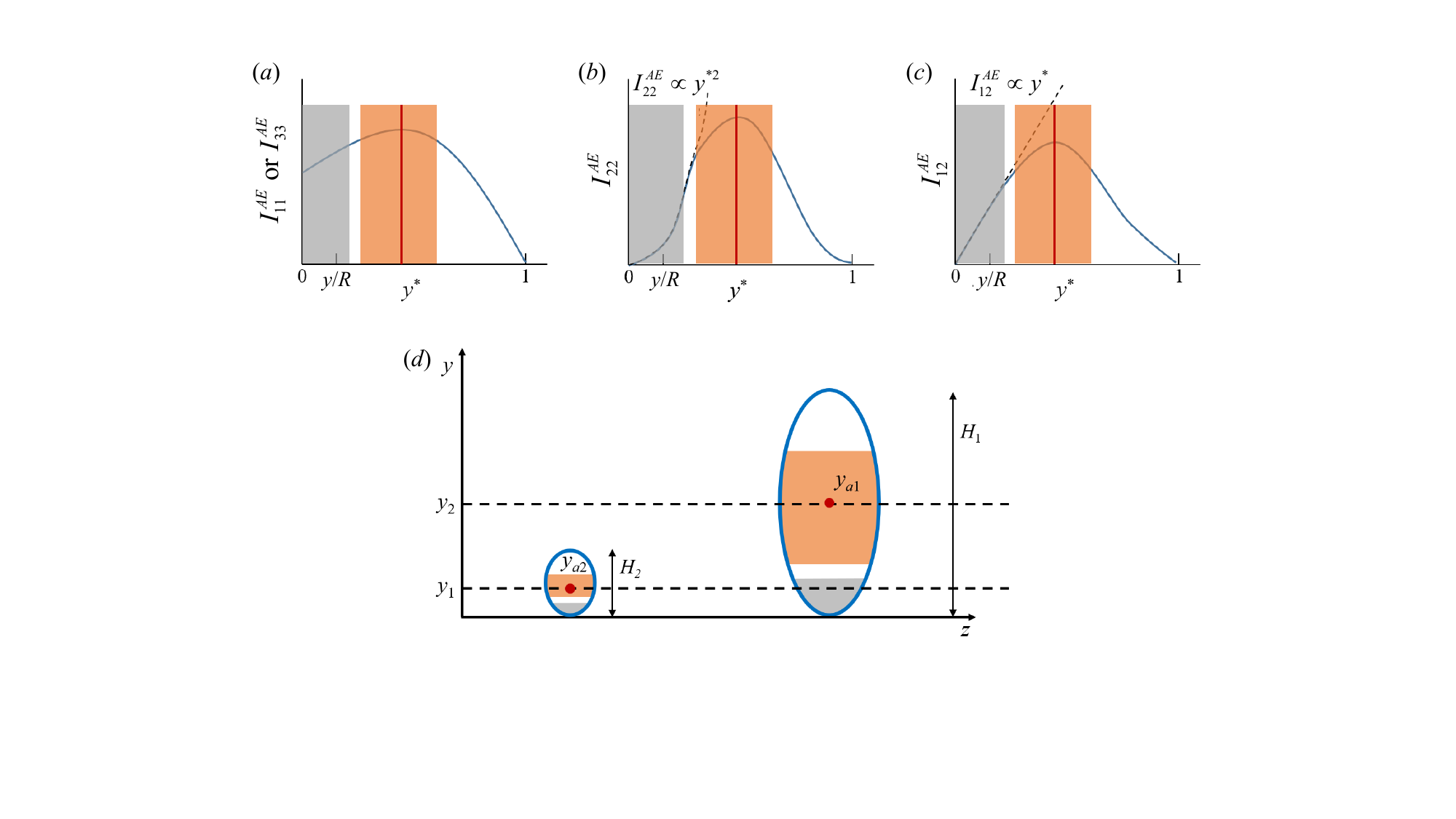}}
\caption{Schematic illustration of the single-eddy statistics and active--inactive contributions of attached eddies (AEs).
($a$--$c$) Single-eddy intensity functions representing the second-order moments, following \citet{Townsend76}: ($a$) streamwise and spanwise components, $I_{11}^{AE}$ and $I_{33}^{AE}$; ($b$) wall-normal component, $I_{22}^{AE}$; and ($c$) Reynolds shear-stress component, $I_{12}^{AE}$.
Here, the profiles are plotted with respect to the eddy-scaled wall-normal location $y^*=y/y_l$.
The red solid lines mark the most active location, $y_a^*$.
($d$) Two AEs of different heights, illustrating that the active or inactive character of their contributions depends on both the observation location and the eddy height.
In ($a$--$d$), the red shaded regions denote the active portions associated with momentum transfer, whereas the grey shaded regions denote the near-wall inactive portions, where the wall-parallel contributions remain finite while the wall-normal and Reynolds shear-stress contributions diminish.}
\label{Fig01}
\end{figure}

Despite the limitations of its inviscid, phenomenological formulation \citep{Marusic19}, Townsend's original idea was further developed into the attached-eddy model (AEM) by Perry and coworkers \citep{Perry82,Perry86}, who introduced geometrically self-similar hierarchies of AEs.
This model provides a unified description not only of the asymptotic behaviour of the turbulence intensities but also of the logarithmic mean-velocity profile and the $k_x^{-1}$ scaling of the streamwise energy spectrum \citep{Perry77}, where $k_x$ is the streamwise wavenumber.
The model was further extended by \citet{Perry86} and \citet{Perry95} to incorporate a broad range of scales in spectral space and the outer wake region.
Moreover, the AEH was also used to explain the growth of the near-wall peak in the streamwise turbulence intensity with Reynolds number \citep{Marusic03}, to model wall-pressure fluctuations \citep{Ahn10}, and to predict the two-point correlations of momentum-transfer and energy-dissipation rates \citep{Mouri17}.
The statistical foundation of the AEM was placed on a more rigorous basis by \citet{Woodcock15}, who applied Campbell's theorem to a random hierarchy of self-similar AEs and showed that velocity moments of arbitrary order and cross-correlations can be derived from the velocity field of a single representative eddy.
These diverse developments share a common foundation in the statistical properties of individual eddies (figure \ref{Fig01}$a$--$c$) conjectured by Townsend.
In this respect, the single-eddy intensity functions constitute the building blocks of the AEM.
However, the intensity functions have typically been constructed from prescribed representative eddies informed by flow visualization observations \citep{Kline67, Head81, Adrian00}.
This naturally raises the question: can the single-eddy intensity functions be extracted directly from turbulent flows?

Obtaining the single-eddy intensity functions is also important because it allows us to link the contributions of individual eddies to the active--inactive concept \citep{Townsend61,Bradshaw67}.
Townsend described active motions as those responsible for turbulent momentum transfer, whereas inactive motions do not contribute to the wall-normal momentum transport (or Reynolds shear stress) at the observation location.
Accordingly, active motions contribute to all three velocity components, whereas inactive motions contribute only to the wall-parallel fluctuations close to the wall, as indicated by the red and grey shaded regions, respectively, in figure~\ref{Fig01}($a$--$c$).
In addition, the intensity functions are assumed to peak at a similar normalized wall-normal location, $y_a^*=y_a/y_l$, within the central active region of the eddy, as indicated by the red solid line.
In this framework, the active contributions at a given observation location are supplied by eddies for which $y=O(y_l)$.
These motions are characterized by $u_\tau$ and $y$, and constitute the universal component of wall turbulence \citep{Townsend61,Bradshaw67}.
In contrast, the inactive contributions arise from larger eddies with heights extending up to the outer scale $R$, and their cumulative superposition leads to the logarithmic variation of the wall-parallel turbulence intensities.
In particular, \cite{Bradshaw67} argued that a principal effect of inactive motion is enhanced viscous dissipation in the viscous sublayer, supplied by turbulent-energy diffusion from the outer layer towards the wall.
Recent studies further demonstrated the relevance of the active--inactive concept. 
\citet{Yang26} found that active motions are associated with the turbulent kinetic energy resolved in two-equation Reynolds-averaged Navier--Stokes models.
In addition, \citet{deGiovanetti16} showed that self-similar AEs in the logarithmic region make a dominant contribution to turbulent skin-friction generation despite their wall-attached portions being inactive in Townsend's sense, while \citet{Deshpande25}, based on the description of inactive motions by \citet{Bradshaw67}, showed that such motions contribute to wall-shear-stress and wall-pressure fluctuations.
It is therefore important to examine active and inactive motions, and the single-eddy intensity functions provide a means of determining how their contributions are distributed within individual eddies, in accordance with Townsend's original concept.

As pointed out by \citet{Hwangy15}, however, active and inactive motions have often been analysed in ways that differ from Townsend's original description.
A key source of this ambiguity is that the active--inactive distinction depends on both the observation location $y$ and the eddy height $y_l$ \citep{Nickels07,Hwangy15,Deshpande21}.
Figure~\ref{Fig01}($d$) illustrates two eddies of heights $H_1$ and $H_2$, with $H_2<H_1$.
At the near-wall observation location $y=y_1$, the smaller eddy of height $H_2$ is active and the larger eddy of height $H_1$ is inactive, whereas at the higher observation location $y=y_2$, the latter can provide an active contribution.
Thus, an individual eddy cannot be classified as entirely active or inactive independently of the observation location.
Distinguishing its active and inactive contributions therefore requires explicit consideration of the wall-normal dependence of the single-eddy intensity functions.

Over the past two decades, extensive experimental \citep{Nickels05,Hultmark12,Marusic13,Vallikivi15,Willert17,Orlu17,Ono23} and numerical efforts \citep{Hoyas06,MK15,Ahn15,Yamamoto18,Pirozzoli21,Hoyas22,Pirozzoli24} have been conducted to examining the statistical predictions of the AEH at high Reynolds numbers and identifying associated self-similar coherent structures \citep{Del06b,Lozano12,Hellstrom16,Hwang18,Yoon20}.
However, comparatively little attention has been paid to extracting the single-eddy intensity functions directly from turbulent flow data.
A novel numerical approach was introduced by \citet{Hwangy15}, who isolated self-sustaining energy-containing motions at prescribed spanwise scales and examined their statistical structure.
These motions exhibited statistical characteristics consistent with the AEH and were subsequently shown to be dynamically self-similar \citep{Hwangy16b}.
However, the resulting motions were obtained through the filtered numerical simulation rather than extracted a posteriori from a fully resolved turbulent flow field.
More recently, \citet{Cheng22} isolated wall-attached eddies of prescribed wall-normal heights using spectral stochastic estimation, although their analysis focused on the streamwise inclination angles rather than the single-eddy intensity functions.
Consequently, the connection between the second-order statistics of motions at a prescribed $y_l$ and the corresponding single-eddy intensity functions, particularly in relation to the active and inactive regions within an individual motion, remains to be fully established.

A different approach is provided by spectral coherence-based analysis, which enables wall-coherent contributions to be isolated from turbulent velocity fields and their statistical characteristics to be examined.
\citet{Baars17} identified self-similar scaling in the coherence between the logarithmic-region turbulence and near-wall velocity fluctuations.
Based on this approach, \citet{Baars20,Baars20b} developed a spectral decomposition to isolate contributions associated with self-similar wall-attached motions and examined their spectral scaling and logarithmic variation of the streamwise turbulence intensity.
\citet{Deshpande21} further used spectral linear stochastic estimation (SLSE) to separate the two-dimensional streamwise energy spectrum into predominantly active and inactive contributions, demonstrating wall scaling of the active component and self-similar characteristics of both components.
In these decompositions, however, the estimated contributions are cumulative over a range of wall-normal coherence heights $y_l$.
Consequently, they provide limited information on how a motion of a prescribed height contributes at different observation locations.

Therefore, the objective of the present study is to extract the single-eddy intensity functions and characterize their dependence on eddy height, paying particular attention to the associated active and inactive regions.
To this end, we employ SLSE to isolate the scale-specific spectral energy contributions associated with wall-coherent motions of prescribed wall-normal coherence height $y_l$, taken here as statistical representations of AEs of corresponding height.
Integration of these spectral contributions over the wall-parallel wavenumbers yields the corresponding single-eddy intensity functions.
Since wall-normal velocity fluctuations are directly associated with momentum-transporting active motions, the single-eddy intensity functions of the streamwise and wall-normal velocity components are compared.
This enables us to address two fundamental questions: (i) where is the active region of an individual wall-coherent motion located, and (ii) where does its inactive contribution begin to emerge towards the wall?
We further examine how the superposition of these contributions relates to the wall-normal variation and Reynolds-number dependence of the streamwise turbulence intensity.
Part 1 addresses these questions by examining the extracted single-eddy intensity functions in the context of the AEH and Townsend's active--inactive description, while Part 2 will analyse the corresponding one- and two-dimensional spectral contributions and their scale-dependent characteristics.

Part 1 is organised as follows.
Section~\ref{sec:classical_AEH} first reviews the classical AEH, focusing on the predicted wall-normal behaviour and the superposition of the single-eddy intensity functions.
Section~\ref{sec:2} then introduces the methodology used to isolate the contribution associated with $y_l$ and to construct the single-eddy intensity functions.
The wall-normal profiles of the resulting intensity functions and their scaling with $y_l$ are examined in \S~\ref{sec:3}, leading to the identification of the active and inactive regions (\S~\ref{sec:3.3}) and the evaluation of their cumulative contributions to the streamwise turbulence intensity (\S\S~\ref{sec:3.4} and \ref{sec:3.5}).
Section~\ref{sec:4} then examines the scaling of the near-wall inactive footprint with $y_l$ and its implications for the Reynolds-number dependence of the near-wall peak, while \S~\ref{sec:5} considers the contributions of wall-coherent and residual motions to the development of the outer peak.
The physical implications of the most active location and the logarithmic variations are further discussed in \S~\ref{sec:6}.




\section{Classical attached-eddy hypothesis}
\label{sec:classical_AEH}
A central feature of the classical AEH is the characteristic wall-normal behaviour of the single-eddy intensity functions, as schematically illustrated in figure~\ref{Fig01}($a$--$c$).
Owing to the self-similar nature of the velocity fields induced by AEs, these functions, $I_{ij}^{AE}(y;y_l)$, depend on $y$ and $y_l$ and can therefore be expressed as $I_{ij}^{AE}(y^*)$ where $y^*=y/y_l$ is the eddy-scaled wall-normal location.
Here, the indices $1$, $2$ and $3$ correspond to the streamwise, wall-normal and spanwise velocity components, respectively, with no summation implied over repeated indices.
In the asymptotic description of the constant-stress layer, viscous effects are neglected, allowing finite slip velocities parallel to the wall while imposing impermeability on the wall-normal velocity.
Consequently, the single-eddy intensity functions for the wall-parallel components, $I_{11}^{AE}$ and $I_{33}^{AE}$, approach finite constants as $y^*\to0$, whereas the wall-normal and Reynolds shear-stress components scale as $I_{22}^{AE}\propto y^{*2}$ and $I_{12}^{AE}\propto y^*$, respectively.
All $I_{ij}^{AE}$ are assumed to vanish for $y^*>1$, corresponding to locations above the eddy height.

The Reynolds stresses can then be represented as a linear superposition of the contributions from AEs of different heights:
\par
\begin{linenomath}
\begin{equation}
\langle u_i u_j\rangle^+(y)
=
\int_y^{y_{l,\max}}
W(y_l) I_{ij}^{AE}(y;y_l)\,
\mathrm{d}y_l,
\label{eq1.1}
\end{equation}
\end{linenomath}
\par\noindent
where $W(y_l)$ denotes the statistical weighting of AEs of height $y_l$, and $y_{l,\max}=O(R)$ is the largest eddy height.
Applying the constant-stress-layer condition gives the inverse-height scaling $W(y_l)\sim 1/y_l$, which describes the population density across the hierarchy of eddy heights \citep{Perry82}.
When $y/y_{l,\max}\ll1$, the streamwise intensity function approaches a finite constant for eddies much taller than the observation location, i.e. $I_{11}^{AE}(y^*)\simeq\mathrm{const.}$ for $y^*\ll1$.
Then, the streamwise turbulence intensity scales as
\par
\begin{linenomath}
\begin{equation}
\langle uu\rangle^+(y)
\sim
\int_y^{y_{l,\max}}
\frac{\mathrm{d}y_l}{y_l}
=
\ln\left(\frac{y_{l,\max}}{y}\right),
\label{eq1.2}
\end{equation}
\end{linenomath}
\par\noindent
where the integral represents the superposition of the inactive contributions corresponding to the shaded region in figure~\ref{Fig01}($a$).
Since $y_{l,\max}=O(R)$, this superposition produces the logarithmic dependence
\par
\begin{linenomath}
\begin{equation}
\langle uu\rangle^+(y)
=
B_1-A_1\ln(y/R),
\label{eq1.3}
\end{equation}
\end{linenomath}
\par\noindent
where $A_1$ and $B_1$ are constants.
An important implication of the inverse-height scaling is that, within the inactive region,
\par
\begin{linenomath}
\begin{equation}
\mathrm{d}\langle uu\rangle^+
\sim
\frac{\mathrm{d}y_l}{y_l}
=
\mathrm{d}\ln(y_l/R).
\label{eq1.4}
\end{equation}
\end{linenomath}
\par\noindent
In other words, each logarithmic interval of eddy height makes an equal incremental contribution to the streamwise turbulence intensity, which is a key feature of the AEM.
Thus, in the classical AEH, the logarithmic variation of the streamwise turbulence intensity arises from the cumulative contribution of the inactive portions of AEs spanning a hierarchy of heights between $y$ and $O(R)$.

By contrast, the active contribution is associated with the region around $y=y_a$, where the wall-normal velocity intensity function remains appreciable.
Here, $y_a$ denotes the wall-normal location at which $I_{22}^{AE}$ attains its maximum.
For self-similar AEs, $y_a$ scales with the eddy height $y_l$, such that $y_a^*=y_a/y_l$ is constant.
At a given observation location $y$, the active contribution is therefore supplied primarily by AEs for which $y_a=O(y)$, whereas the contributions from much taller AEs diminish because $I_{22}^{AE}\propto y^{*2}$ as $y^*\to0$.
Consequently, the wall-normal turbulence intensity does not accumulate over a broad hierarchy of eddy heights but approaches a constant:
\par
\begin{linenomath}
\begin{equation}
\langle vv\rangle^+(y)=B_2,
\label{eq1.5}
\end{equation}
\end{linenomath}
\par\noindent
where $B_2$ is a constant.
This scale-local behaviour distinguishes the active contribution from its cumulative inactive contribution of the wall-parallel components.

Note that \citet{Perry82}, who formulated a hierarchical distribution of AEs, referred to the single-eddy intensity functions as ``eddy hierarchy functions'' and illustrated their superposition schematically in figures 21 and 22 of their study.
The active--inactive interpretation of this superposition was also re-highlighted in \citet{Deshpande25}.
In the following section, we examine the wall-normal characteristics of the single-eddy intensity functions extracted using the present method in the context of the classical AEH.

\section{Pipe-flow DNS dataset and methodology for isolating wall-coherent motions}
\label{sec:2}

\subsection{Pipe-flow DNS dataset}
\label{sec:2.1}

\begin{table}
\centering
\begin{tabular}{cccccccccc}
\hline
$Re_{b}$ & $Re_{\tau}$ & $L_x/R$ & $(N_x, N_r, N_\theta)$ & $\Delta x^+$ & $\Delta (R\theta)^+$ & $\Delta r_{\min}^+$ & $\Delta r_{\max}^+$ & $\Delta t^+$ & Plotting style\\
\hline
35\,000   & 930   & $10\pi$  & (4097, 301, 1025) & 6.84 & 5.73 & 0.334 & 9.24 & 0.25 & \plotstyle{black}{\fullcirc} \\
133\,000   & 3008  & $30$   & (12289, 901, 3073) & 7.34 & 6.15 & 0.361  & 9.91 & 0.20 & \plotstyle{ReBlue}{\fullcirc} \\
280\,000   & 5933  & $3\pi$  & (6145, 973, 6145) & 9.10 & 6.07 & 0.050 & 8.15 & 0.16  & \plotstyle{ReRed}{\fullcirc} \\
\hline
\end{tabular}
\caption{Simulation parameters. Here, $Re_{b}$ and $Re_{\tau}$ is the bulk mean Reynolds number and the friction Reynolds number, respectively. $L_x$ is the domain size in the streamwise direction, and $R$ indicates the pipe radius. $N_x$, $N_r$, and $N_\theta$ indicate the number of grid points in the streamwise, radial, and azimuthal directions, respectively. The streamwise and spanwise grid sizes are denoted by $\Delta x^+$ and $\Delta (R\theta)^+$, respectively. The minimum and maximum radial grid spacings are represented by $\Delta r_{\mathrm{min}}^+$ and $\Delta r_{\mathrm{max}}^+$, respectively, and $\Delta t^+$ is the time step.}
\label{tab:Table1}
\end{table}

The DNS dataset of the fully-developed turbulent pipe flows at $Re_\tau \approx$ $930$ \citep{Ahn13,Lee15,Hwang16b}, $3000$ \citep{Ahn15}, and $6000$ \citep{Kim26} was used in the present work.
Here, $Re_\tau=u_\tau R/\nu$ is the friction Reynolds number, where $\nu$ is the kinematic viscosity and $R$ is the pipe radius.
Throughout this paper, $R$ also denotes the outer length scale, such as channel half-height or boundary layer thickness.
The incompressible Navier--Stokes equations were solved in cylindrical coordinates using a fraction step method \citep{Kim02, Jang11}.
The time advancement was performed using the Crank--Nicolson method, and the spatial derivatives were discretized using a second-order central difference scheme on a staggered grid.
The uniform grids were employed in the streamwise ($x$) and azimuthal ($\theta$) directions, whereas a non-uniform grid was used in the radial ($r$) direction.
The wall-normal direction is defined as $y = R - r$, where $R$ is the pipe radius.
The periodic boundary conditions were imposed in the $x$ and $\theta$ directions, and the no-slip condition was applied at the pipe wall.
At the coordinate singularity on the pipe centreline, the radial velocity was obtained by averaging the values at diametrically opposite points \citep{Jang11}.
The simulation parameters are summarized in table \ref{tab:Table1}, and further numerical details are provided in the aforementioned studies.
For $Re_\tau = 6000$, the streamwise domain length was $L_x/R = 3\pi$, which is shorter than the $L_x/R = 30$ employed for the other two cases.
Nevertheless, this domain length is sufficiently long to capture the wall-attached motions considered in the present study.
As demonstrated later, the active contribution associated with individual AEs begins to decay for $y_l/R \gtrsim 0.5$.
A similar cutoff is observed in the lower-Reynolds-number cases with $L_x/R\simeq30$, suggesting that the decay is not caused by the shorter domain and supporting the adequacy of the $Re_\tau=6000$ domain for the present analysis.
Further support is provided by \citet{Kim26}, whose one-point statistics at $Re_\tau=6000$ agree well with those from the longer-domain simulation ($L_x/R=15$) of \citet{Pirozzoli21}.
Note that the simulation of $Re_\tau = 6000$ was performed on the NURION system at the Korea Institute of Science and Technology Information (KISTI), using the open-source library PaScaL\_TDMA \citep{Kim21}\footnote{Available at \url{https://github.com/MPMC-Lab/PaScaL_TDMA}.}, which reduces the communication overhead associated with distributed tridiagonal solvers on massively parallel architectures.
A total of 1230, 688 and 1712 instantaneous flow fields were used for the cases at $Re_\tau\approx930$, 3000 and 6000, respectively. 
The corresponding sampling periods exceeded 13\,000 viscous time units for all cases and 40\,000 viscous time units for the $Re_\tau\approx6000$ case. 
Over the range primarily considered here, $y_l/R\lesssim0.5$, these sampling periods correspond to approximately ten or more characteristic eddy-turnover times, estimated as $T_e\sim y_l/u_\tau$, and are therefore considered sufficient for statistical convergence.

In the present study, the streamwise, wall-normal, and azimuthal velocity fluctuations are denoted by $u$, $v$, and $w$, respectively.
Ensemble averages are denoted by angled brackets $\langle \cdot \rangle$, and the superscript $+$ indicates normalization in viscous units based on the friction velocity $u_\tau$ and the viscous length scale $\delta_\nu = \nu/u_\tau$.

\subsection{Spectral linear stochastic estimation}
\label{sec:2.2}
To isolate wall-coherent motions associated with a prescribed wall-normal coherence height $y_l$, we employ two-point spectral linear stochastic estimation (SLSE) in the two-dimensional streamwise--azimuthal wavenumber space \citep{Mad19,Deshpande21}.
Linear stochastic estimation (LSE) was introduced by \citet{Adrian79} as a minimum-mean-square-error approach for estimating a conditional velocity field associated with a given event from unconditional flow statistics.
More generally, the conditioning event can be defined using signals measured at different spatial locations and can encompass a prescribed range of event amplitudes \citep{Adrian96}.
This approach can also be formulated in the spectral domain \citep{Tinney06}, where a scale-dependent linear transfer function is obtained from the cross-spectral relationships between the conditioning and target signals.
Unlike conventional single-time LSE, this spectral formulation retains the scale-dependent amplitude and phase relationships between signals measured at different locations.
In wall-bounded turbulence, SLSE was used to isolate the contribution of the streamwise energy spectrum that is coherent across wall-normal locations \citep{Baars17,Mad19,Baars20}.

For the streamwise velocity fluctuations, the two-dimensional Fourier coefficient at $y=y_1$ can be linearly estimated from that at a different wall-normal location $y=y_2$ through a transfer function $H_u(y_1,y_2;k_x, k_\theta)$,
\par
\begin{linenomath}
\begin{equation}
\hat{u}'(y_1;k_x, k_\theta) 
= 
H_u(y_1,y_2;k_x, k_\theta) 
\hat{u}(y_2;k_x, k_\theta),
\end{equation}
\end{linenomath}
\par\noindent
where $\hat{\cdot}$ denotes the two-dimensional Fourier coefficient, and the superscript prime indicates an estimated quantity.
In the present work, $k_x(=2\pi/\lambda_x)$ and $k_\theta(=2\pi /\lambda_\theta)$ denote the streamwise and azimuthal wavenumbers, respectively, with $\lambda_x$ and $\lambda_\theta$ indicating the corresponding wavelengths.
The linear transfer function $H_u(y_1,y_2;k_x, k_\theta)$ is given by
\par
\begin{linenomath}
\begin{equation}
H_u(y_1,y_2;k_x, k_\theta) 
= 
\frac{\langle \hat{u}(y_1;k_x, k_\theta)\hat{u}^{*}(y_2;k_x, k_\theta) \rangle}
{\langle \hat{u}(y_2;k_x, k_\theta)\hat{u}^*(y_2;k_x, k_\theta) \rangle},
\end{equation}
\end{linenomath}
\par\noindent
Here, the superscript asterisk denotes the complex conjugate.
The numerator is the cross-spectrum between the velocity signals at $y_1$ and $y_2$, whereas the denominator is the auto-spectrum at $y_2$.
The transfer function is generally complex and can be written as $H_u=|H_u|\exp(\mathrm{i}\psi_u)$, where $\psi_u$ represents the scale-dependent phase difference between the velocity signals at the two wall-normal locations.

Accordingly, the two-dimensional spectral contribution at $y_1$ associated with the velocity fluctuations at $y_2$ is defined as
\par
\begin{linenomath}
\begin{equation}
\phi_{uu}^{\,\mathrm{2D}}
(y_1|y_2;k_x,k_\theta)
=
|H_u(y_1,y_2;k_x,k_\theta)|^2
\phi_{uu}^{\,\mathrm{2D}}
(y_2;k_x,k_\theta).
\label{eq2.3}
\end{equation}
\end{linenomath}
\par\noindent
Here, $y_2$ is referred to as the reference location, while $y_1=y_o$ denotes the observation location.
Thus, $\phi_{uu}^{\,\mathrm{2D}}(y_o|y_2)$ represents the spectral energy at $y_o$ that is linearly coherent with the velocity fluctuations at $y_2$.
In the present analysis, however, $|H_u|^2$ in \eqref{eq2.3} is replaced by $[\Re(H_u)]^2$, and the notation $\phi_{uu}^{\,\mathrm{2D}}(y_o|y_2)$ is retained for the spectral contribution evaluated using the in-phase component of the transfer function.
The physical motivation for this modification and its quantitative influence on the extracted intensity functions are examined in \S~\ref{sec:2.4}.

\subsection{Isolation of motions associated with a prescribed wall-normal coherence height}
\label{sec:2.3}
Following the linear-superposition representation of the classical AEH in \eqref{eq1.1}, we assume that the spectral contributions associated with different ranges of wall-normal coherence height are additively separable \citep{Townsend76,Perry82}.
Under this assumption, contributions over prescribed ranges of coherence height can be isolated by differencing cumulative spectra estimated using reference locations at different wall-normal positions, following previous spectral-coherence-based decomposition frameworks \citep{Baars17,Baars20,Deshpande21}.
In the present study, this differencing procedure is applied to the modified spectral contribution obtained by replacing $|H_u|^2$ in \eqref{eq2.3} with $[\Re(H_u)]^2$, as described in \S~\ref{sec:2.2}.
Importantly, the limitation of the cumulative decomposition discussed below arises from its treatment of a range of wall-normal coherence heights and is therefore independent of whether the spectral contribution is evaluated using $|H_u|^2$ or $[\Re(H_u)]^2$.

\begin{figure}
\centerline{\includegraphics[trim=0.2cm 10.0cm 0.5cm 0.2cm,width=14cm]{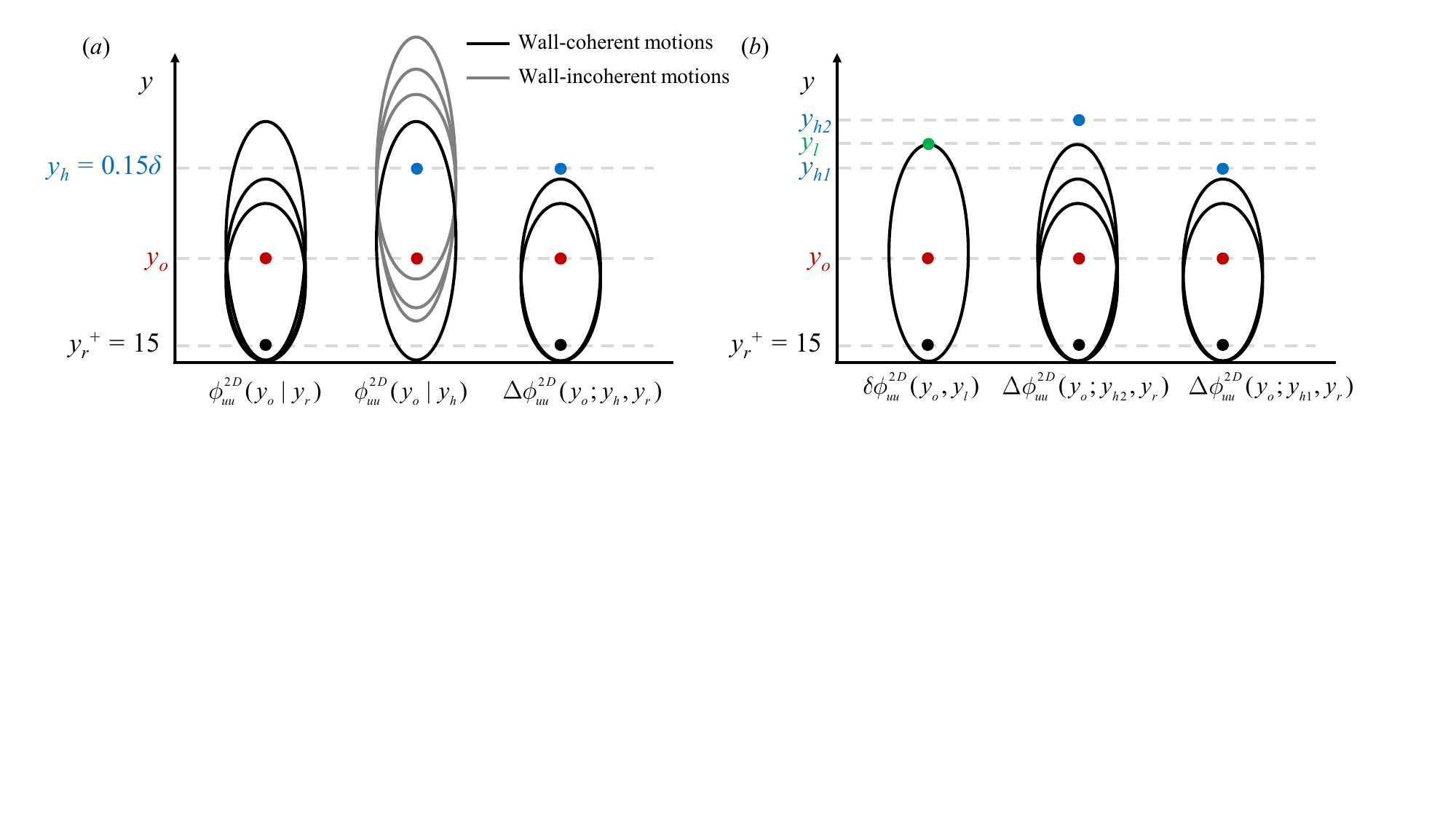}}
 \caption{Schematic illustration of the motions contributing to the spectral estimates and their decomposition.
The ellipses denote motions whose velocity signatures encompass the marked wall-normal locations; the black and grey ellipses represent wall-coherent and wall-incoherent motions, respectively.
The black, blue and red dots mark the near-wall reference location ($y_r^+=15$), the off-wall reference locations and the observation location ($y_o$), respectively.
($a$) Isolation of the cumulative spectral contribution from wall-coherent motions with $y_o\leq y_l\leq y_h$, denoted by $\Delta\phi_{uu}^{\,\mathrm{2D}}(y_o;y_h,y_r)$.
($b$) Isolation of the single-eddy spectral contribution $\delta\phi_{uu}^{\,\mathrm{2D}}(y_o;y_l)$ associated with the narrow coherence-height interval between $y_{h_1}$ and $y_{h_2}$, represented by $y_l=(y_{h_1}+y_{h_2})/2$.}
 \label{Fig02}
\end{figure}

Figure~\ref{Fig02} schematically illustrates the eddies contributing to the estimated spectral energy at $y_o$ for different choices of the reference location. 
The ellipses represent eddies whose velocity signatures contribute to the corresponding estimated or decomposed spectra.
In particular, figure~\ref{Fig02}($a$) represents the cumulative spectral decomposition employed by
\citet{Deshpande21}. 
Two reference locations are used to obtain separate estimates of the spectrum at $y_o$, and their difference isolates the contribution from wall-coherent motions whose $y_l$ lie within a certain range.
The leftmost schematic in figure~\ref{Fig02}($a$) corresponds to the near-wall reference location, $y_2^+=y_r^+=15$. 
Since only motions whose velocity signatures encompass both $y_r$ and $y_o$ contribute to the estimated spectrum, $\phi_{uu}^{\,\mathrm{2D}}(y_o|y_r)$ represents the cumulative spectral contribution at $y_o$ from wall-coherent motions with $y_l\gtrsim y_o$.
Thus, when $y_o$ lies within the logarithmic region, this cumulative contribution also includes wall-attached outer-scaled motions with $y_l=O(R)$, which are associated with very-large-scale motions \citep{Kim99} or global modes \citep{Del03}.
Here, wall-coherent motions are defined as those whose velocity signatures exhibit linear coherence with the near-wall reference signal at $y_r^+=15$; motions lacking such coherence are referred to as wall-incoherent motions.

The second estimate is obtained using an off-wall reference location, $y_2=y_h$, positioned above $y_o$, as illustrated by the middle schematic in figure~\ref{Fig02}($a$).
In this case, $\phi_{uu}^{\,\mathrm{2D}}(y_o|y_h)$ contains contributions from motions whose velocity signatures extend across both $y_o$ and $y_h$.
Since coherence with the near-wall reference location, $y_r$, is not imposed in this estimate, it can contain both wall-coherent and wall-incoherent motions, represented by the black and grey ellipses, respectively.
In \citet{Deshpande21}, the off-wall reference location was chosen as $y_h=0.15R$, corresponding to the upper bound of the logarithmic region.
For $y_o \ll y_h$, this estimate can therefore be interpreted as the cumulative spectral contribution from tall motions extending across the two locations.

The positive part of the difference between $\phi_{uu}^{\,\mathrm{2D}}(y_o|y_r)$ and $\phi_{uu}^{\,\mathrm{2D}}(y_o|y_h)$ is defined as
\par
\begin{linenomath}
\begin{equation}
\Delta\phi_{uu}^{\,\mathrm{2D}}
(y_o;y_h,y_r)
=
\max\left\{
\phi_{uu}^{\,\mathrm{2D}}(y_o|y_r)
-
\phi_{uu}^{\,\mathrm{2D}}(y_o|y_h),
\,0
\right\}.
\label{eq2.4}
\end{equation}
\end{linenomath}
\par\noindent
In this operation, the contributions from tall wall-coherent motions with $y_l\gtrsim y_h$, which are contained in both estimated spectra, are removed by subtraction. 
Wall-incoherent motions contained only in $\phi_{uu}^{\,\mathrm{2D}}(y_o|y_h)$ yield negative differences and are excluded by retaining only the positive part. 
Consequently, $\Delta\phi_{uu}^{\,\mathrm{2D}}$ represents the cumulative spectral contribution from wall-coherent motions in the range $y_o\le y_l\le y_h$, as illustrated in the rightmost schematic in figure~\ref{Fig02}($a$).

Although \citet{Deshpande21} interpreted this quantity as representing inactive contributions from AEs, it is intrinsically cumulative over the range $y_o\leq y_l\leq y_h$.
Because this range extends down to $y_l=O(y_o)$, $\Delta\phi_{uu}^{\,\mathrm{2D}}$ can contain contributions from wall-coherent motions that are active at $y_o$, even when $y_o\ll y_h$.
As $y_o$ is moved progressively farther below $y_h$, the range of included coherence heights also broadens, incorporating progressively smaller wall-coherent motions that are expected to be active at $y_o$.
\citet{Deshpande21} also interpreted the residual spectrum,
$\phi_{uu}^{\,\mathrm{2D}}(y_o)-\phi_{uu}^{\,\mathrm{2D}}(y_o|y_r)$,
as being predominantly associated with active motions.
However, this residual comprises wall-incoherent contributions spanning a broad range of scales and therefore carries no explicit information about $y_l$.
Thus, to examine Townsend's active--inactive concept at the level of an individual AE, $y_l$ should be held fixed while the observation location $y_o$ is varied in the wall-normal direction.

Our approach is as follows. 
Figure~\ref{Fig02}($b$) extends the cumulative decomposition in figure~\ref{Fig02}($a$) by allowing the off-wall reference location $y_{h_1}=y_h$ to vary in the wall-normal direction, rather than fixing it at the upper bound of the logarithmic region.
The resulting spectrum, $\Delta\phi_{uu}^{\,\mathrm{2D}}(y_o;y_{h_1},y_r)$, is illustrated in the rightmost schematic in figure~\ref{Fig02}($b$).
This decomposition is determined by three wall-normal locations: the near-wall reference location $y_r$, the observation location $y_o$ and the off-wall reference location $y_{h_1}$.
Here, $y_r$ defines the distinction between wall-coherent and wall-incoherent motions, whereas $y_o$ and $y_{h_1}$ define the cumulative range $y_o\leq y_l\leq y_{h_1}$.

We then introduce a second off-wall reference location immediately above $y_{h_1}$, denoted by $y_{h_2}=y_{h_1}+\Delta y$, and construct $\Delta\phi_{uu}^{\,\mathrm{2D}}(y_o;y_{h_2},y_r)$.
This spectrum is cumulative over $y_o\leq y_l\leq y_{h_2}$.
It therefore contains the contribution represented by $\Delta\phi_{uu}^{\,\mathrm{2D}}(y_o;y_{h_1},y_r)$ as a subset, which is illustrated in the middle schematic in figure~\ref{Fig02}($b$).
Consequently, the difference between the two cumulative spectra isolates the spectral contribution associated with the narrow interval $y_{h_1}\leq y_l\leq y_{h_2}$:
\par
\begin{linenomath}
\begin{equation}
\delta\phi_{uu}^{\,\mathrm{2D}}
(y_o;y_l)
=
\Delta\phi_{uu}^{\,\mathrm{2D}}
(y_o;y_{h_2},y_r)
-
\Delta\phi_{uu}^{\,\mathrm{2D}}
(y_o;y_{h_1},y_r),
\label{eq2.5}
\end{equation}
\end{linenomath}
\par\noindent
where $y_l=(y_{h_1}+y_{h_2})/2$ is taken as the representative coherence height of the finite interval $[y_{h_1},y_{h_2}]$.
Note that the positive-part operation in \eqref{eq2.4} is essential to the present construction.
Without it, $\phi_{uu}^{\,\mathrm{2D}}(y_o|y_r)$ would cancel identically in \eqref{eq2.5}, leaving $\delta\phi_{uu}^{2D} = \phi_{uu}^{\,\mathrm{2D}}(y_o|y_{h_1})-\phi_{uu}^{\,\mathrm{2D}}(y_o|y_{h_2})$, which no longer carries any information on wall coherence.
The clipping in \eqref{eq2.4} retains the dependence on $y_r$, and thereby ensures that $\delta\phi_{uu}^{2D}$ represents the contribution from wall-coherent motions only.

In general, the two off-wall reference locations can be separated by a larger wall-normal distance to isolate the cumulative spectral contribution over a prescribed range of coherence heights, $y_{\min}\leq y_l\leq y_{\max}$.
We denote this contribution by $\Delta\phi_{uu}^{\,\mathrm{2D}}(y_o;y_{\min}\leq y_l\leq y_{\max})$, with the dependence on the fixed near-wall reference location $y_r^+=15$ omitted for brevity.
Here, $\Delta\phi_{uu}^{\,\mathrm{2D}}$ denotes the cumulative contribution over a finite range of $y_l$, whereas $\delta\phi_{uu}^{\,\mathrm{2D}}$ denotes the contribution associated with an individual coherence-height band centred at $y_l$.
If the prescribed range is partitioned into $M$ adjacent bands centred at $y_{l,m}$, the following identity holds:
\par
\begin{linenomath}
\begin{equation}
\Delta\phi_{uu}^{\,\mathrm{2D}}
(y_o;y_{\min}\leq y_l\leq y_{\max})
=
\sum_{m=1}^{M}
\delta\phi_{uu}^{\,\mathrm{2D}}
(y_o;y_{l,m}).
\label{eq2.6}
\end{equation}
\end{linenomath}
\par\noindent

In the present analysis, $y_{h_1}$ and $y_{h_2}$ used to evaluate each $\delta\phi_{uu}^{\,\mathrm{2D}}$ are chosen as two adjacent wall-normal grid points, such that the resulting contribution corresponds to a narrow coherence-height band centred at $y_l$.
With $y_l$ fixed in this manner, the observation location $y_o$ is varied in the wall-normal direction and is hereafter denoted simply by $y$.

Then, by integrating the spectral contribution in \eqref{eq2.5} over the wall-parallel wavenumbers, we can obtain the wall-normal profile of the streamwise turbulence intensity associated with a wall-coherent motion of $y_l$ as
\par
\begin{linenomath}
\begin{equation}
I_{uu}(y;y_l)
=
\frac{1}{\Delta y_l}
\int_0^\infty\int_0^\infty
\delta\phi_{uu}^{\,\mathrm{2D}}
(y;y_l;k_x,k_\theta)
\,\mathrm{d}k_x\,\mathrm{d}k_\theta .
\label{eq2.7}
\end{equation}
\end{linenomath}
\par\noindent
Here, $\Delta y_l=y_{h_2}-y_{h_1}=\Delta y$ is the width of the finite coherence-height interval.
Since the magnitude of the extracted contribution depends on $\Delta y_l$, the integrated spectral contribution is normalized by this interval width.
As a result, $I_{uu}(y;y_l)$ represents the contribution to the streamwise velocity variance at $y$ per unit coherence-height interval centred at $y_l$.
By varying $y$ while holding $y_l$ fixed, $I_{uu}(y;y_l)$ describes the wall-normal profile of the contribution from a wall-coherent motion with $y_l$ and is hereafter referred to as the single-eddy intensity function.
Within the classical AEH, the extracted $I_{uu}(y;y_l)$ corresponds to the data-derived counterpart of the integrand $W(y_l)I_{11}^{AE}(y;y_l)$ in \eqref{eq1.1}, i.e. the statistically weighted contribution per unit eddy-height interval from AEs of height $y_l$.
The same procedure is applied to the wall-normal velocity fluctuations:
\par
\begin{linenomath}
\begin{equation}
I_{vv}(y;y_l)
=
\frac{1}{\Delta y_l}
\int_0^\infty\int_0^\infty
\delta\phi_{vv}^{\,\mathrm{2D}}
(y;y_l;k_x,k_\theta)
\,\mathrm{d}k_x\,\mathrm{d}k_\theta .
\label{eq2.8}
\end{equation}
\end{linenomath}
\par\noindent
We focus on $I_{uu}$ and $I_{vv}$ because their comparison provides a basis for interpreting the extracted contributions within Townsend's active--inactive framework \citep{Townsend61}.
The present SLSE analysis treats the streamwise and wall-normal velocity components separately using single-input transfer functions.
A multiple-input formulation that jointly incorporates multiple velocity components \citep[e.g.,][]{Sasaki19,Ying26} could explicitly account for cross-component coherence and is left for future work.

\subsection{Physical interpretation and validation of the single-eddy decomposition}
\label{sec:2.4}
In this section, we examine the physical interpretation and internal consistency of the single-eddy decomposition, focusing on the use of $[\Re(H_{u_j})]^2$ rather than $|H_{u_j}|^2$, where $u_1=u$ and $u_2=v$.
Because the active--inactive interpretation requires the streamwise and wall-normal contributions to be compared at the same eddy-scaled observation location, $y^*=y/y_l$ (figure~\ref{Fig01}), the phase sensitivity retained by the real-component treatment provides a more direct basis for comparing their co-located wall-normal footprints.
We therefore examine the spatial organization of the $u$- and $v$-velocity signatures, quantify the influence of the phase treatment on the resulting single-eddy intensity functions, and assess the numerical consistency of the decomposition.

To examine the spatial organization relevant to the present phase treatment, we consider the normalized two-point velocity correlation, defined as
\par
\begin{linenomath}
\begin{equation}
R_{u_j u_j}(r_x,y,r_z;y_{\mathrm{ref}})
=
\frac{
\left\langle
u_j(x+r_x,y,z+r_z)
u_j(x,y_{\mathrm{ref}},z)
\right\rangle
}{
\sqrt{
\left\langle u_j^2(y)\right\rangle
\left\langle u_j^2(y_{\mathrm{ref}})\right\rangle}
}.
\label{eq2.9}
\end{equation}
\end{linenomath}
\par
\begin{figure}
\centerline{\includegraphics[trim=2.5cm 1.0cm 2.0cm 0.2cm,width=14cm]{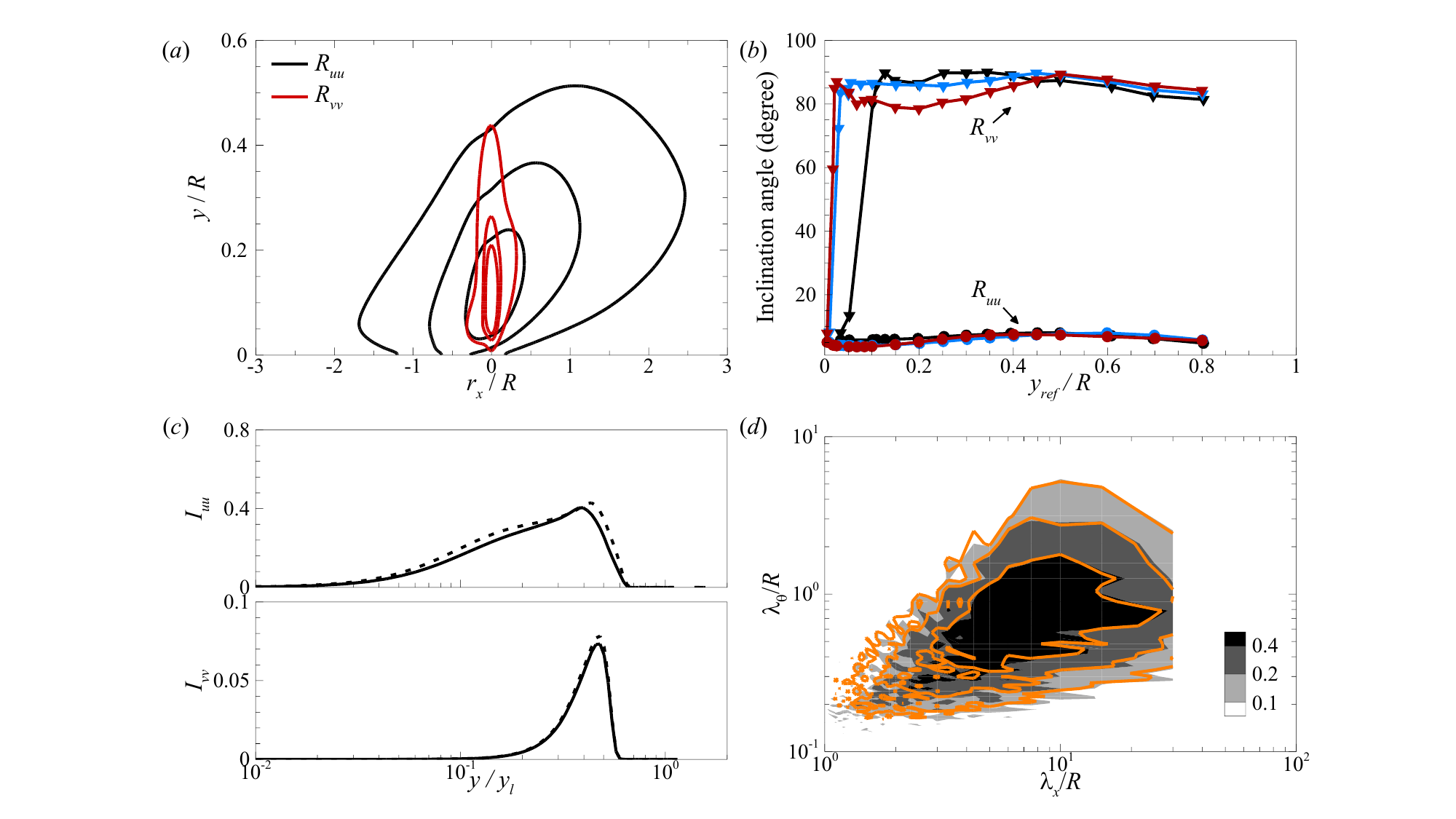}}
 \caption{($a$) Two-point correlations of the streamwise and wall-normal velocity fluctuations, $R_{uu}$ and $R_{vv}$, respectively, in the $x$--$y$ plane ($r_z=0$) at $y_{\mathrm{ref}}=0.1R$ for $Re_\tau=930$.
The black and red contours denote $R_{uu}$ and $R_{vv}$, respectively; the contour levels are $R_{uu}=0.15$, $0.25$, and $0.4$, and $R_{vv}=0.05$, $0.15$, and $0.25$.
($b$) Inclination angles of $R_{uu}$ (circles) and $R_{vv}$ (triangles), evaluated from the $0.25$ correlation contours, as functions of $y_{\mathrm{ref}}/R$ for $Re_\tau=930$ (black), $3000$ (blue), and $6000$ (red).
($c$) Single-eddy intensity functions $I_{uu}$ and $I_{vv}$ for $y_l=0.1R$ at $Re_\tau=930$.
The solid and dashed lines denote the estimates obtained using $[\Re(H_{u_j})]^2$ and $|H_{u_j}|^2$, respectively.
($d$) Comparison of $\sum \delta\phi_{uu}^{\,\mathrm{2D}}(y;y_l)$ over $0.1R\leq y_l\leq0.3R$ (filled contours) with the corresponding cumulative contribution, $\Delta\phi_{uu}^{\,\mathrm{2D}}(y;0.1R\leq y_l\leq0.3R)$ (line contours), at $y=0.08R$ and $Re_\tau=930$.
Both spectra are premultiplied by $k_xk_\theta$, and the contours are shown at $0.1$, $0.2$, and $0.4$ of their respective maxima.}
 \label{Fig03}
\end{figure}

\noindent
Here, $z=R\theta$ denotes the wall-based azimuthal coordinate, and $y_{\mathrm{ref}}$ denotes the reference height used only for the present correlation analysis.
Figure~\ref{Fig03}($a$) shows $R_{uu}$ (black) and $R_{vv}$ (red) in the $x$--$y$ plane ($r_z=0$) at $y_{\mathrm{ref}}=0.1R$ for $Re_\tau=930$.
As seen, the contours of $R_{uu}$ are elongated in the streamwise direction and extend towards the wall, consistent with the inclined nature of large-scale $u$ structures \citep{Liu01, Adrian07}.
In contrast, $R_{vv}$ has a much smaller streamwise extent and is predominantly elongated in the wall-normal direction, resulting in a substantially larger inclination angle than that of $R_{uu}$ \citep{Liu01,Sillero14}.

Although these two-point correlations contain contributions from a broad range of scales and should not be interpreted as representing an individual eddy, their low-level contours provide a statistical measure of the mean spatial organization of the velocity signatures associated with the reference location \citep{Baltzer13,Hwang16b,Hwang16}.
At the correlation level of 0.25, the upper extent of the $R_{uu}$ contour at $r_x=0$ reaches $y/R\simeq0.3$, whereas the $R_{vv}$ contour is confined primarily to $0.05\lesssim y/R\lesssim0.2$.
The latter range may indicate where appreciable streamwise and wall-normal velocity signatures coexist, while below this range the wall-normal signature rapidly diminishes.
The nearly wall-normal orientation of $R_{vv}$ further indicates that its coherence across wall-normal locations involves little streamwise displacement.
Accordingly, the use of $[\Re(H_{u_j})]^2$ places greater weight on contributions with little relative phase displacement, providing a more direct basis for comparing the co-located wall-normal footprints of $u$ and $v$ at the same observation location.

To quantify this difference, figure~\ref{Fig03}($b$) shows the inclination angles of $R_{uu}$ and $R_{vv}$ as functions of $y_{\mathrm{ref}}$.
The inclination angle is evaluated at a fixed correlation level of $0.25$ using the ellipse-based definition of \citet{Sillero14}, in which an equivalent ellipse is constructed from the correlation contour and the angle is measured between its major semiaxis and the positive $x$ direction.
The inclination angle of $R_{uu}$ remains approximately $10^\circ$ over a substantial range of $y_{\mathrm{ref}}$, whereas that of $R_{vv}$ is generally greater than $70^\circ$.
In particular, the inclination angle of $R_{vv}$ increases rapidly for $y_{\mathrm{ref}}^+\gtrsim100$ and subsequently remains close to $90^\circ$.
These trends are consistently observed at all three Reynolds numbers considered here and agree with the two-point correlation characteristics reported for turbulent channels and boundary layers by \citet{Sillero14}.
Although the quantitative values depend to some extent on the selected correlation level, the marked difference between the orientations of $R_{uu}$ and $R_{vv}$ persists across the contour levels examined.

This difference in the inclination angles has an important implication for the SLSE-based decomposition.
The nearly vertical orientation of $R_{vv}$ implies little streamwise displacement of the wall-normal velocity signature between different wall-normal locations, whereas the inclined $R_{uu}$ signature is associated with a finite streamwise displacement.
In Fourier space, such a displacement appears as a scale-dependent phase difference in the transfer function $H_{u_j}$.
While $|H_{u_j}|$ retains the coherent amplitude irrespective of this phase difference, $\Re(H_{u_j})$ represents its in-phase component at the same streamwise position.
Accordingly, the use of $[\Re(H_{u_j})]^2$ places greater weight on contributions with little relative phase displacement, providing a more direct basis for comparing the co-located wall-normal footprints of $u$ and $v$ and thereby identifying the active and inactive regions of an individual wall-coherent motion.
Note that previous studies used the cross-spectral phase to determine the streamwise displacement of individual Fourier modes and thereby estimate the scale-dependent inclination angles of wall-attached motions \citep{Deshpande19,Cheng22}.
Extending the present coherence-height decomposition to retain phase information at a prescribed $y_l$ could reveal whether the streamwise inclination of individual wall-coherent motions varies systematically with their coherence height, although such an analysis is beyond the scope of the present study.

To quantify the influence of the phase treatment on the extracted single-eddy intensity, figure~\ref{Fig03}($c$) compares $I_{uu}$ and $I_{vv}$ at $y_l=0.1R$ obtained using the real-component estimate, $[\Re(H_{u_j})]^2$, with those obtained using the magnitude-based estimate, $|H_{u_j}|^2$.
The solid and dashed lines denote the former and latter estimates, respectively.
For $I_{uu}$, the two estimates exhibit a small but discernible difference: the estimate based on $|H_u|^2$ is slightly larger in the lower and upper portions of the wall-coherent motion and exhibits a modest outward shift of the peak location.
Nevertheless, the ratio of the intensity obtained using $[\Re(H_u)]^2$ to that obtained using $|H_u|^2$ remains above approximately 0.9 across the wall-normal direction, indicating that the real-component estimate retains most of the energetic contribution.
By contrast, the two estimates of $I_{vv}$ are nearly indistinguishable, with essentially no change in the peak location.
This contrast is consistent with the correlation patterns in figure~\ref{Fig03}($a$,$b$): the nearly vertical orientation of $R_{vv}$ involves little relative streamwise displacement between wall-normal locations and hence weak sensitivity to the phase treatment, whereas the inclined $R_{uu}$ signature involves a finite phase displacement.
Although a small difference remains for $I_{uu}$, its magnitude is minor compared with the overall intensity.
Thus, the use of the real-component estimate has only a limited influence on the energetic magnitude of the extracted contributions while weighting them towards their in-phase, co-located components.
Similar trends are observed for other values of $y_l$ and $Re_\tau$, although these results are omitted for brevity.

Finally, the consistency of the proposed extraction procedure is examined using the reconstruction relation in \eqref{eq2.6}, as shown in figure~\ref{Fig03}($d$).
Although this relation follows algebraically from the band-wise definition in \eqref{eq2.5}, the comparison provides a check that the cumulative contribution is consistently recovered from the individual coherence-height bands in the discrete implementation, including the pointwise positive-part operation in \eqref{eq2.4}.
Here, the coherence-height range $0.1R\leq y_l\leq0.3R$ is considered.
The line contours represent the cumulative spectral contribution over the entire range, corresponding to the left-hand side of \eqref{eq2.6}, whereas the filled contours represent the sum of the individual coherence-height-band contributions on the right-hand side.
The comparison is made at $y=0.08R$, close to the peak of the corresponding cumulative intensity profile.
As shown, the two spectra exhibit nearly overlapping energetic regions and contour shapes in the wall-parallel wavelength space.
This agreement confirms that the cumulative contribution over the prescribed range of $y_l$ is consistently recovered by recombining the individual coherence-height bands.
Similar agreement is observed at other wall-normal locations and Reynolds numbers, although these results are omitted for brevity.
The spectral characteristics of the single-eddy contributions are examined in detail in Part~2.

\section{Single-eddy intensity functions and their scaling}
\label{sec:3}
In this section, we examine the wall-normal profiles of the single-eddy intensity functions, $I_{uu}$ and $I_{vv}$, defined in \eqref{eq2.7} and \eqref{eq2.8}, respectively, in the context of the classical AEH.
We first assess whether their profile shapes exhibit self-similarity when expressed in terms of the eddy-scaled wall-normal coordinate, $y^*=y/y_l$, and whether they satisfy the boundary behaviour expected for AEs, as illustrated schematically in figure~\ref{Fig01}.
We then examine the scaling of their amplitudes with $y_l$ and the Reynolds-number dependence of the scaled profiles.
Finally, based on the contrasting wall-normal characteristics of $I_{uu}$ and $I_{vv}$, we identify the active and inactive regions associated with an individual wall-coherent motion.

\subsection{Self-similarity of the single-eddy intensity functions}
\label{sec:3.1}
\begin{figure}
\centerline{\includegraphics[trim=2.5cm 10.0cm 2.5cm 25.0cm,width=14cm]{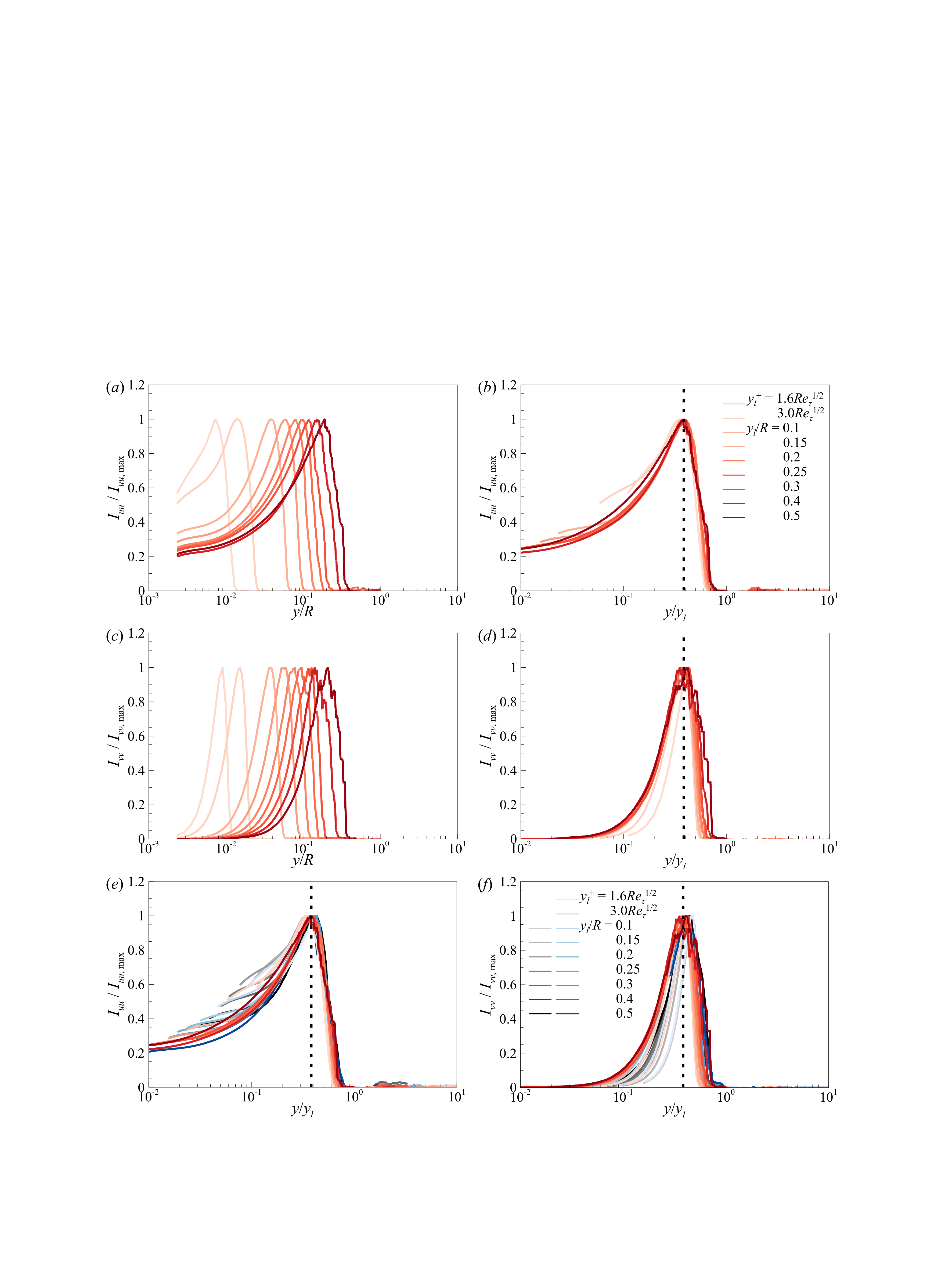}}
 \caption{The wall-normal profiles of the single-eddy intensity functions, $I_{uu}$ ($a$,$b$) and $I_{vv}$ ($c$,$d$) at $Re_\tau=6000$.
 The profiles are plotted against ($a$,$c$) the outer-scaled wall-normal location, $y/R$, and ($b$,$d$) the eddy-scaled wall-normal location, $y^*=y/y_l$.
 Comparison of ($e$) $I_{uu}$ and ($f$) $I_{vv}$ as functions of $y^*$ for $Re_\tau=930$ (black), $3000$ (blue), and $6000$ (red).
 For each set of profiles, the colour shade changes from light to dark with increasing $y_l$.
 All profiles are normalized by their respective maxima.
 Here, the inserted vertical dashed line denotes $y^*=0.39$.}
 \label{Fig04}
\end{figure}

Figure~\ref{Fig04}($a$) shows $I_{uu}$ for different values of $y_l$ at $Re_\tau=6000$.
The intensity functions are plotted against the outer-scaled wall-normal location, $y/R$, with each profile normalized by its respective maximum.
The colour shade changes from light to dark with increasing $y_l$, which spans the range from $y_l^+=1.6Re_\tau^{1/2}$ to $y_l/R=0.5$.
The lower bound is chosen to coincide with the location of the maximum Reynolds shear stress, $y_m^+\simeq1.6Re_\tau^{1/2}$ \citep{Klewicki21}; below this height, the extracted intensity functions become increasingly influenced by near-wall viscous effects across $y$.
As $y_l$ increases, the peak of $I_{uu}$ progressively shifts away from the wall.
When the wall-normal coordinate is normalized by $y_l$, however, the profiles exhibit a good collapse, as shown in figure~\ref{Fig04}($b$).
In particular, their peaks align near $y^*\simeq0.39$, as indicated by the vertical dashed line, demonstrating that the peak location scales with $y_l$.
The profiles are also consistent with the boundary behaviour expected for an AE (figure~\ref{Fig01}): the intensity becomes negligible for $y^*\gtrsim1$, whereas it remains finite towards the wall.
Because the near-wall reference location is fixed at $y_r^+=15$ and only data for $y^+\geq15$ are shown, progressively smaller values of $y^*$ are accessible as $y_l$ increases.
As a result, the finite near-wall behaviour is most evident for the larger values of $y_l$, for which the normalized intensity approaches approximately $I_{uu}/I_{uu,\max}\simeq0.2$.

Next, we examine the wall-normal component $I_{vv}$ shown in figure~\ref{Fig04}($c$,$d$).
Similar to $I_{uu}$, the profiles progressively shift away from the wall with increasing $y_l$ when plotted against $y/R$ (figure~\ref{Fig04}$c$), whereas they exhibit a reasonable collapse when expressed as functions of $y^*$ (figure~\ref{Fig04}$d$).
The profiles decay rapidly to zero as $y^*$ approaches unity and attain their maxima at $y^*\simeq0.38$--$0.40$, consistent with the peak location of $I_{uu}$ in figure~\ref{Fig04}($b$).
We denote this common peak location by $y_a^*$ and refer to it as the most active location of an individual wall-coherent motion, because the wall-normal velocity contribution associated with active motions is strongest.
The region surrounding $y_a^*$ can therefore be associated with the active region described by \citet{Townsend61}, characterized here by an appreciable wall-normal velocity contribution.
Notably, $y_a^*$ lies below the geometric centre of the motion, $y^*=0.5$, and is numerically close to the commonly reported value of the von Kármán constant, $\kappa\simeq0.39$ \citep{Marusic13}.
The possible physical significance of this correspondence is discussed further in \S~\ref{sec:6.1}.

Toward the wall, $I_{vv}$ decreases rapidly for $y^*\lesssim0.10$--$0.15$ and becomes very small, whereas $I_{uu}$ remains finite, as shown in figure~\ref{Fig04}($b$,$d$).
The rapid reduction in $I_{vv}$ near this range can therefore be interpreted as a transition from the active region surrounding $y^*=y_a^*$ to the lower inactive portion of the motion, where the streamwise velocity contribution remains finite but the wall-normal contribution becomes weak.
Such behaviour is consistent with Townsend's active--inactive description \citep{Townsend61,Bradshaw67} that large AEs contribute to the Reynolds shear stress farther from the wall while appearing inactive at a lower observation location.
The present intensity functions provide an eddy-scaled representation of this picture, demonstrating that an individual eddy is not intrinsically active or inactive; rather, the nature of its contribution depends on the wall-normal observation location, as illustrated schematically in figure~\ref{Fig01}.

To examine the Reynolds-number dependence of the trends described above, figure~\ref{Fig04}($e$,$f$) compares the normalized $I_{uu}$ and $I_{vv}$ profiles, respectively, at $Re_\tau=930$ (black), $3000$ (blue), and $6000$ (red).
In general, the profiles exhibit a reasonable collapse when expressed as functions of $y^*$.
For both $I_{uu}$ and $I_{vv}$, the peak locations remain within $y_a^*\simeq0.38$--$0.40$ over the Reynolds numbers considered, with particularly good agreement near the most active location.
Similarly, the rapid reduction of $I_{vv}$ begins within the range $0.10\lesssim y^*\lesssim0.15$, consistent with the active-to-inactive transition identified above.

For relatively small $y_l$, $I_{vv}$ decreases more rapidly toward the wall, as seen in figure~\ref{Fig04}($d$,$f$).
At a given $y^*$, a smaller $y_l^+$ corresponds to a smaller wall-normal location, $y^+=y^*y_l^+$, such that the lower portion of the wall-coherent motion enters the viscous near-wall region at a comparatively larger value of $y^*$.
This effect is evident for the smallest coherence height at $Re_\tau=6000$ (e.g., $y_l^+=1.6Re_\tau^{1/2}\approx124$) and becomes more pronounced at $Re_\tau=930$, where even $y_l/R=0.1$ corresponds to $y_l^+<100$.
In addition, the fixed near-wall reference location, $y_r^+=15$, limits the accessible lower extent of each profile to $y^*_{\min}=15/y_l^+$; resolving a profile down to $y^*=0.1$ therefore requires $y_l^+\gtrsim150$.
Consistent with these considerations, the lower portions of the $I_{vv}$ profiles exhibit a reasonable collapse once $y_l^+$ exceeds approximately $200$.
Thus, despite the stronger near-wall influence at smaller $y_l^+$, the principal features identified above, including the most active location near $y_a^*\simeq0.38$--$0.40$ and the reduction of the wall-normal contribution toward the inactive region, are consistently observed across the Reynolds numbers.

\subsection{Hierarchical scaling of the single-eddy intensity}
\label{sec:3.2}

\begin{figure}
\centerline{\includegraphics[trim=2.5cm 1.0cm 2.5cm 0.2cm,width=14cm]{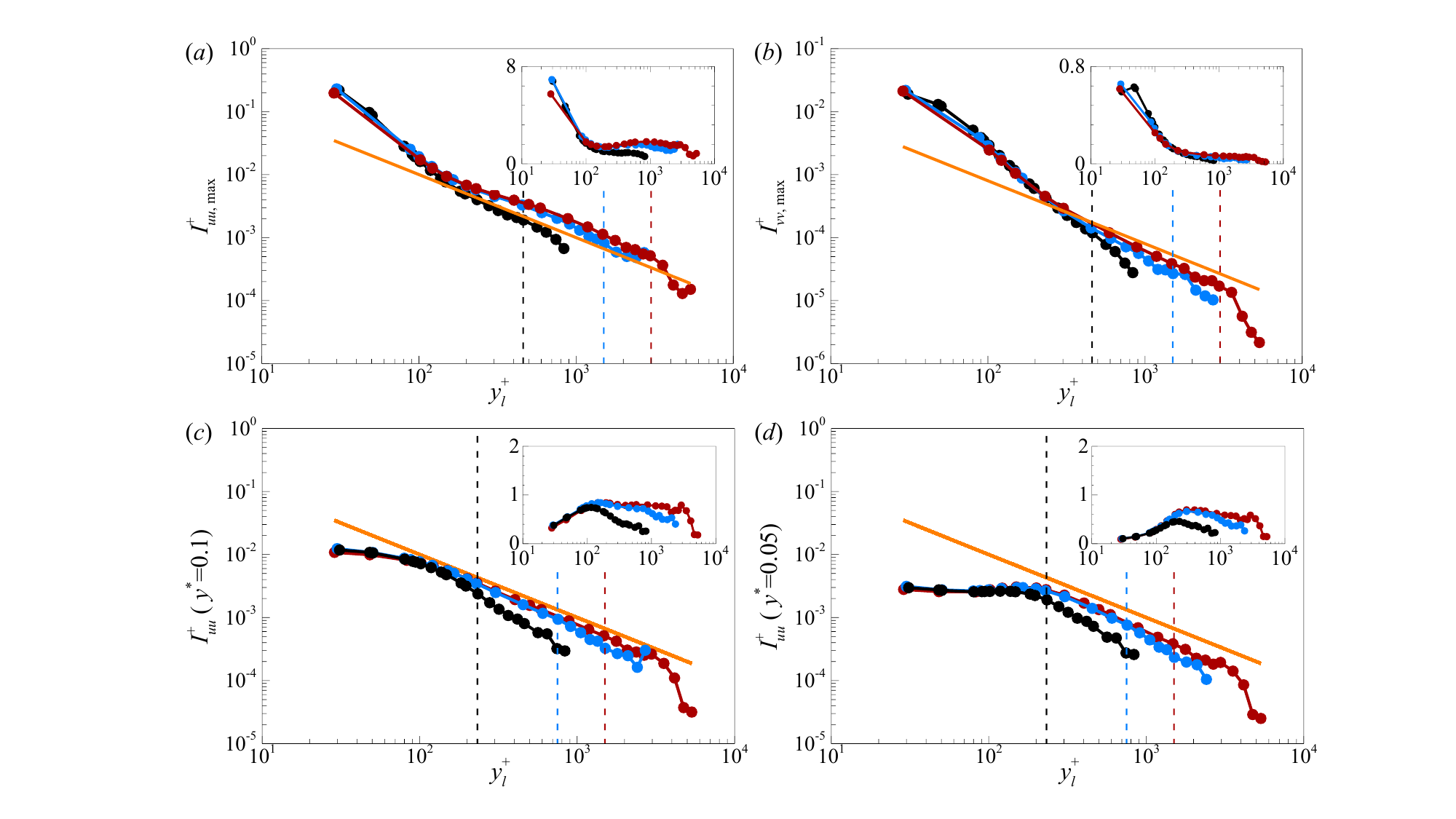}}
 \caption{The peak values of the single-eddy intensity functions at $y^*=y_a^*$ as a function of $y_l^+$: ($a$) $I_{uu,\max}^+$ and ($b$) $I_{vv,\max}^+$. 
The variations of $I_{uu}^+$ at $y^*=0.1$ and $0.05$ are shown in ($c$) and ($d$), respectively.
In ($a$,$b$), the vertical lines denote $y_l/R=0.5$, whereas those in ($c$,$d$) denote $y_l/R=0.25$. 
The solid orange lines indicate the $(y_l^+)^{-1}$ scaling. 
The insets show the corresponding premultiplied intensities, $y_l^+I_{uu}^+$ in ($a$,$c$,$d$) and $y_l^+I_{vv}^+$ in ($b$).
The coherence height spans the range from $y_l^+=30$ to $y_l/R=0.9$. 
The colour coding is as specified in table~\ref{tab:Table1}.}
 \label{Fig05}
\end{figure}

We next examine how the peak values of the single-eddy intensity functions vary with $y_l$ to determine the scaling of the intensity at the most active location ($y=y_a$).
Figure~\ref{Fig05}($a$,$b$) shows $I_{uu,\max}^+$ and $I_{vv,\max}^+$ as functions of $y_l^+$.
For both components, the data at different Reynolds numbers exhibit a reasonable collapse over a broad range of $y_l^+$.
Interestingly, for $y_l^+\gtrsim150$--$200$, both $I_{uu,\max}^+$ and $I_{vv,\max}^+$ decrease approximately following a $(y_l^+)^{-1}$ scaling, as indicated by the solid orange lines, before exhibiting a rapid cutoff for $y_l/R\gtrsim0.5$ (denoted by the vertical dashed lines).
This scaling is shown more clearly in the insets, where the premultiplied peak intensities, $y_l^+I_{uu,\max}^+$ and $y_l^+I_{vv,\max}^+$, exhibit approximately constant plateaus.

Given that the peak intensities in figure~\ref{Fig05}($a$,$b$) are expressed in viscous units, the observed collapse is consistent with $u_\tau$ being the characteristic velocity scale of the wall-coherent motions for both the streamwise and wall-normal components.
Together with the collapse of the normalized profiles in figure~\ref{Fig04}, this inverse-height dependence supports a self-similar hierarchy in both profile shape and amplitude over the approximate range $200\lesssim y_l^+\lesssim0.5R^+$.
In addition, the observed inverse scaling implies
\par
\begin{linenomath}
\begin{equation}
\mathrm{d}\langle u_j u_j\rangle_{y_a}^+
=I_{u_ju_j}^+(y=y_a)\,\mathrm{d}y_l^+
\sim
\frac{1}{y_l^+}\,\mathrm{d}y_l^+
=\mathrm{d}\ln y_l^+,
\label{eq3.1}
\end{equation}
\end{linenomath}
\par\noindent
where the subscript $y_a$ indicates that each incremental contribution is evaluated at the most active location, $y_a=y_a^*y_l$, corresponding to that coherence height.
Thus, equal logarithmic intervals of $y_l^+$ provide approximately equal incremental contributions when evaluated at the respective most active locations. 
Equivalently,
\par
\begin{linenomath}
\begin{equation}
\frac{\mathrm{d}\langle u_j u_j\rangle_{y_a}^+}
{\mathrm{d}\ln y_l^+}
\approx \mathrm{constant}.
\label{eq3.2}
\end{equation}
\end{linenomath}
\par\noindent
These relations have the same equal-contribution-per-logarithmic-interval form as the hierarchical scaling of the inactive contribution in \eqref{eq1.4}.
Whereas \eqref{eq1.4} describes the accumulation of inactive streamwise contributions at a fixed observation location, \eqref{eq3.1} and \eqref{eq3.2} describe the contributions of individual coherence-height bands at their respective most active locations.
As discussed in \S~\ref{sec:2.3}, the extracted $I_{u_ju_j}(y;y_l)$ corresponds, in the context of the classical AEH, to the data-derived counterpart of the weighted integrand $W(y_l)I_{jj}^{AE}(y;y_l)$ in \eqref{eq1.1}.
Thus, the inverse-height scaling observed for both $I_{uu,\max}$ and $I_{vv,\max}$ demonstrates that this weighted contribution retains the hierarchical $y_l^{-1}$ scaling in the active region as well.

To examine whether the inverse-height scaling persists toward the lower portion of the wall-coherent motions, figures~\ref{Fig05}($c$,$d$) show $I_{uu}^+$ as a function of $y_l^+$ at fixed $y^*=0.1$ and $0.05$, respectively.
As observed in figures~\ref{Fig04}($e$,$f$), these locations lie within the inactive portion of the wall-coherent motion, where $I_{vv}$ becomes weak while $I_{uu}$ remains finite.
As at the most active location, $I_{uu}^+$ follows an approximate $(y_l^+)^{-1}$ scaling at these lower $y^*$.
The scaling range is more limited, however, extending approximately over $200\lesssim y_l^+\lesssim0.25R^+$, beyond which $I_{uu}^+$ departs from the inverse-height behaviour.
This range is more clearly identified from the premultiplied representations in the insets: for $Re_\tau=3000$ and $6000$, $y_l^+I_{uu}^+$ exhibits an approximately constant plateau before decreasing for $y_l/R\gtrsim0.25$.
No comparable plateau is evident at $Re_\tau=930$, because the constraint $y_l^+\geq15/y^*$ imposed by the near-wall reference location leaves insufficient separation from the outer-scale cutoff.
The inverse-height scaling in the inactive portion therefore becomes discernible only when sufficient scale separation exists between the viscous lower limit and the outer-scale cutoff.
The earlier departure from this scaling, compared with the cutoff near $y_l/R\simeq0.5$ at the most active location, further suggests that the inactive footprint is more sensitive to finite outer-scale effects than the active portion of the wall-coherent motion.

The observed inverse-height scaling provides evidence for the uniform weighting per logarithmic interval of eddy height underlying hierarchical AEM.
Whereas this property is prescribed through the continuous hierarchy in \citet{Perry82}, the present $I_{u_ju_j,\max}^+\sim(y_l^+)^{-1}$ scaling emerges directly from the weighted single-eddy intensity functions extracted from turbulent flow, without prescribing their dependence on $y_l$ a priori.
The departure from the inverse-height scaling at large $y_l/R$ also can be interpreted in the context of outer-scale modifications to the AE hierarchy.
\citet{Perry86,Perry95} introduced such modifications to account for the distinct behaviour of the largest eddies as their sizes approach the outer length scale.
Consistent with this picture, \citet{Hwang18} found that wall-attached streamwise structures extracted in instantaneous flow fields follow an inverse-height population distribution over the self-similar range, whereas the tallest structures depart from this distribution and lose geometric self-similarity \citep[see also][]{Hwang20}.
Similarly, \citet{Yoon20} distinguished self-similar attached motions from outer-scaled, non-self-similar wall-attached structures as their heights approach the boundary-layer thickness.
In this respect, the cutoff near $y_l/R\simeq0.5$ at the most active location may therefore reflect a transition from the self-similar attached hierarchy to outer-scaled motions as $y_l$ becomes $O(R)$.

In the inactive portion, the inverse-height scaling breaks down earlier, near $y_l/R\simeq0.25$, indicating that the lower footprint of a large wall-coherent motion is more sensitive to departures from ideal self-similarity.
\citet{Hwangy16} showed that the wall-parallel velocity fluctuations associated with logarithmic- and outer-layer motions are influenced by viscous wall effects below their peak locations, leading to a partial breakdown of self-similarity in the inactive portion.
In addition, \citet{Hwang20} showed that outer-scaled, non-self-similar wall-attached structures can contaminate the self-similar behaviour expected in the logarithmic layer.
More recently, \citet{Hwangy24} demonstrated that, at fixed near-wall locations, the outer-scaled spectral energy associated with large-scale motions of size $O(R)$ decreases with increasing $Re_\tau$, indicating a Reynolds-number-dependent attenuation of their near-wall footprint.
A closely related picture was reported by \citet{Pirozzoli24}, who showed that the near-wall signature of AEs becomes progressively weaker as their centres (analogous to $y_a$ in the present work) move away from the wall and attributed this attenuation to viscous effects associated with turbulent Stokes layers.
Although these studies do not correspond exactly to the present comparison at fixed $y^*$, they collectively indicate that the lower footprint of a wall-attached motion is particularly susceptible to both viscous and outer-scale effects.
This interpretation is consistent with the earlier cutoff found here in the inactive portion than at the most active location.

\subsection{Active and inactive regions of the single-eddy intensity}
\label{sec:3.3}
\begin{figure}
\centerline{\includegraphics[trim=2.5cm 9.0cm 2.5cm 0.2cm,width=14cm]{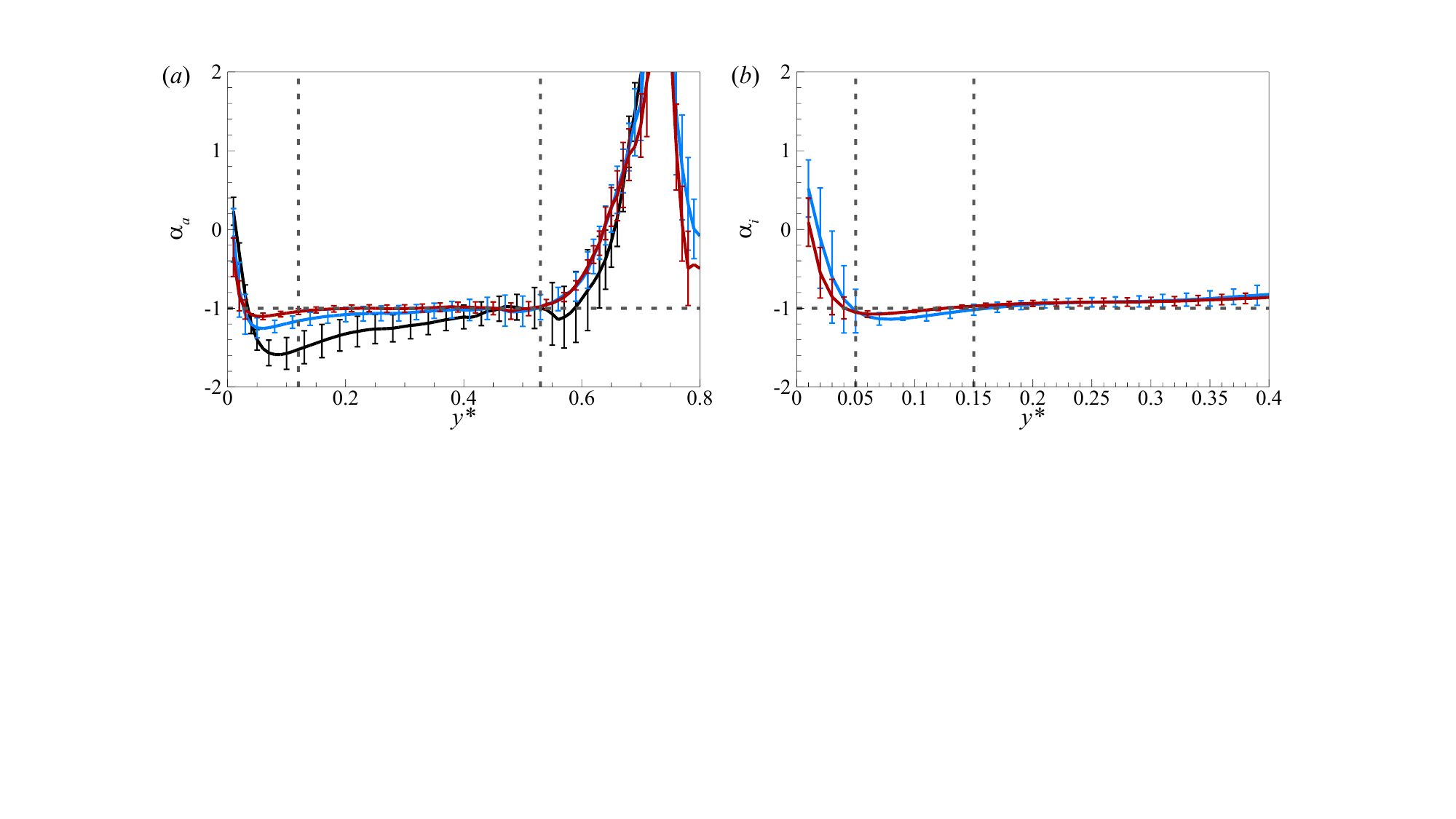}}
 \caption{Fitted exponent $\alpha$ of $I_{uu}^+\sim(y_l^+)^\alpha$ as a function of $y^*$, obtained over ($a$) $200\leq y_l^+\leq0.5R^+$ and ($b$) $200 \leq y_l^+ \leq 0.25R^+$. 
 The error bars indicate the 95\% confidence intervals. 
 The horizontal dashed line denotes $\alpha=-1$. 
 The vertical dashed lines indicate the approximate ranges over which $\alpha\approx-1$: $0.12 \leq y^* \leq 0.53$ in ($a$) and $0.05 \leq y^* \leq 0.15$ in ($b$).
 The colour coding is as specified in table~\ref{tab:Table1}.}
 \label{Fig06}
\end{figure}

The inverse-height scaling identified in figure~\ref{Fig05} is now considered across the eddy-scaled wall-normal coordinate ($y^*$) to determine how its extent relates to the active and inactive portions of the wall-coherent motions.
The coherence-height range over which this behaviour is observed depends on the relative wall-normal location within the motion.
At the most active location, it extends over $200\lesssim y_l^+\lesssim0.5R^+$, whereas at the lower locations, $y^*=0.1$ and $0.05$, the corresponding range is restricted to $200\lesssim y_l^+\lesssim0.25R^+$.
To account for this difference, power-law fits are performed over the respective ranges to characterize the active and inactive portions in figures~\ref{Fig06}($a$) and ($b$).

We first consider the broader scaling range identified near the most active location.
As shown in figure~\ref{Fig04}($e$,$f$), the normalized $I_{uu}$ and $I_{vv}$ profiles exhibit particularly good agreement near $y^*=y_a^*$, where the wall-normal contribution reaches its maximum.
To determine the extent over which the inverse-height scaling persists around this active region, $I_{uu}^+$ is fitted at each fixed $y^*$ according to $I_{uu}^+\sim(y_l^+)^{\alpha_a}$ over $200\leq y_l^+\leq0.5R^+$.
Figure~\ref{Fig06}($a$) shows the resulting exponent $\alpha_a$ as a function of $y^*$.
For $Re_\tau=3000$ and $6000$, a distinct region with $\alpha_a\approx-1$ is observed over $0.12\lesssim y^*\lesssim0.53$, as indicated by the vertical dashed lines.
This interval contains the most active location, $y_a^*\simeq0.38$--$0.40$, and its lower bound is consistent with the active-to-inactive transition identified from the reduction of $I_{vv}$ in \S~\ref{sec:3.1}.
At $Re_\tau=930$, the available scale separation is more limited, and a similarly extended plateau in $\alpha_a$ is not evident; nevertheless, $\alpha_a$ remains close to $-1$ from the vicinity of $y^*=y_a^*$ up to $y^*\simeq0.53$.

To examine the scaling in the lower inactive portion, the same analysis is repeated using the more restricted coherence-height range identified in figure~\ref{Fig05}($c$,$d$).
At each fixed $y^*$, $I_{uu}^+$ is fitted according to $I_{uu}^+\sim(y_l^+)^{\alpha_i}$ over $200\leq y_l^+\leq0.25R^+$, and figure~\ref{Fig06}($b$) shows the resulting exponent $\alpha_i$.
The $Re_\tau=930$ case is excluded because the available scale separation is insufficient to obtain a reliable fit over this range.
For $Re_\tau=3000$ and $6000$, $\alpha_i$ remains close to $-1$ over $0.05\lesssim y^*\lesssim0.15$.
Although the inverse-height scaling may extend to even smaller $y^*$, the uncertainty of the fitted exponent increases substantially for $y^*<0.05$, as indicated by the error bars.
Thus, when the Reynolds number is sufficiently high to provide adequate scale separation, the $(y_l^+)^{-1}$ scaling is not restricted to the active portion of the wall-coherent motions but also persists into the inactive region close to the wall, recovering the classical inactive scaling described by \eqref{eq1.4}.

Based on these observations, we define the active and inactive regions in terms of the eddy-scaled wall-normal location $y^*$ and the coherence height $y_l$.
Specifically,
\par
\begin{linenomath}
\begin{subequations}
\label{eq3.3}
\begin{align}
\left.
\begin{aligned}
0.15 &\lesssim y^* \lesssim 0.5,\\
200 &\lesssim y_l^+ \lesssim 0.5R^+
\end{aligned}
\right\}
&\qquad \text{active portion},
\label{eq3.3a}\\[3pt]
\left.
\begin{aligned}
y^* &\lesssim 0.1,\\
200 &\lesssim y_l^+ \lesssim 0.25R^+
\end{aligned}
\right\}
&\qquad \text{inactive portion}.
\label{eq3.3b}
\end{align}
\end{subequations}
\end{linenomath}
\par\noindent
Here, the ranges in $y^*$ define the active and inactive portions of an individual wall-coherent motion, while the ranges in $y_l^+$ indicate the coherence-height intervals over which the $(y_l^+)^{-1}$ scaling is observed.
The onset of the inactive portion $y^*\simeq0.1$ also corresponds to the observation in figure~\ref{Fig04}($e$,$f$), where $I_{vv}$ falls below approximately $5\%$ of its maximum while $I_{uu}$ remains finite at approximately $20\%$ of its maximum.
Since the reduction of $I_{vv}$ occurs gradually over $y^*\simeq0.1$--$0.15$, this boundary should be regarded as a transition between the active and inactive portions.
The range $y^*\gtrsim0.5$--$0.55$ marks an upper decay transition from the active inverse-height scaling range, beyond which both intensity functions rapidly decrease and approach zero as $y^*\to1$.
For the inactive portion, the condition $0.25R^+\gtrsim200$, or equivalently $Re_\tau\gtrsim800$, is required for an inverse-height range to exist.
This range broadens with Reynolds number and reaches nearly one decade at $Re_\tau=6000$.
Although the ranges in \eqref{eq3.3} require further validation over a wider range of Reynolds numbers, they are adopted here as practical criteria for distinguishing the active and inactive portions of an individual wall-coherent motion.

\subsection{Active and inactive contributions to the streamwise turbulence intensity}
\label{sec:3.4}
We now examine the cumulative contributions of the active and inactive portions of wall-coherent motions using the regions defined in \eqref{eq3.3}.
Within the classical AEH discussed in \S~\ref{sec:classical_AEH}, the cumulative contribution from the active portions is expected to remain approximately constant across the constant-stress layer, whereas the accumulation of the inactive portions gives rise to the logarithmic variation of the streamwise turbulence intensity.
We therefore examine whether these characteristic behaviours can be recovered directly by integrating $I_{uu}(y;y_l)$ over the ranges of coherence height corresponding to the active and inactive portions at each observation location.

\begin{figure}
\centerline{\includegraphics[trim=2.5cm 1.0cm 2.5cm 0.2cm,width=14cm]{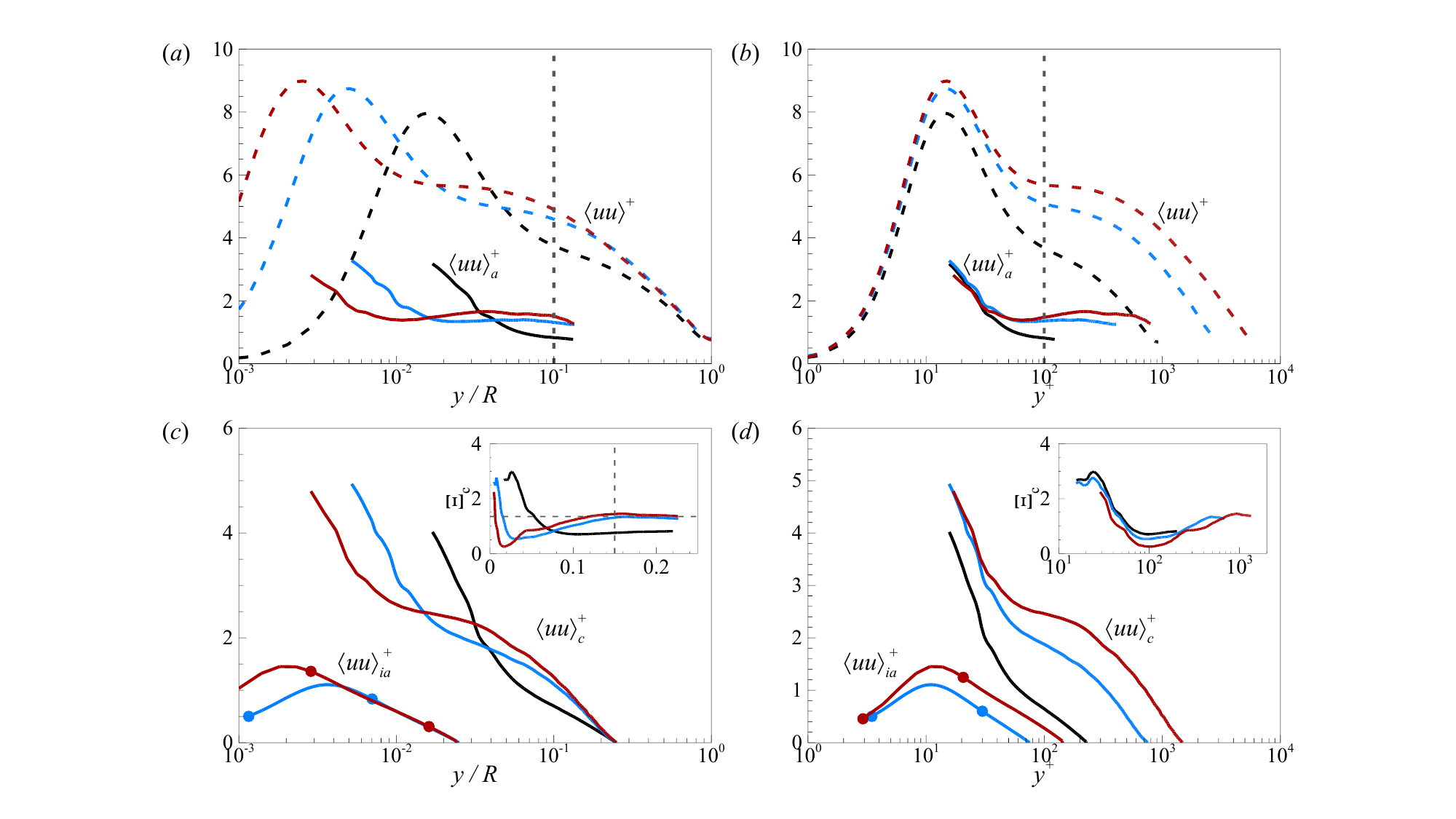}}
 \caption{Cumulative contributions of wall-coherent motions to the streamwise turbulence intensity. 
 ($a$,$b$) Active contribution, $\langle uu\rangle_a^+$, obtained by integrating over $0.15 \le y^* \le 0.5$ (solid lines). 
 For comparison, the total streamwise turbulence intensity, $\langle uu\rangle^+$, is shown by the dashed lines. 
 ($c$,$d$) Inactive contribution, $\langle uu\rangle_{ia}^+$, obtained by integrating over $10y \le y_l \le 0.25R$ (solid lines with circles), together with the cumulative contribution, $\langle uu\rangle_c^+$, obtained over $2y \le y_l \le 0.5R$ (solid lines). 
 The insets in ($c$,$d$) show the logarithmic indicator function of $\langle uu\rangle_c^+$, $\Xi_c=-y^+\partial\langle uu\rangle_c^+/\partial y^+$.
 The wall-normal coordinate is plotted in outer units $y/R$ ($a$,$c$), and in inner units $y^+$ ($b$,$d$).
 The colour coding is as specified in table~\ref{tab:Table1}.}
 \label{Fig07}
\end{figure}

Figure~\ref{Fig07}($a$,$b$) shows the cumulative active contribution, $\langle uu\rangle_a^+$, obtained by integrating $I_{uu}$ over the $y^*$ range in \eqref{eq3.3a}.
Since $y^*=y/y_l$, this range corresponds, for a given $y$, to $2y\leq y_l\leq y/0.15$.
Then, the active contribution can be evaluated as
\par
\begin{linenomath}
\begin{equation}
\langle uu \rangle_a^+(y^+)
=\int_{2y^+}^{y^+/0.15}
I_{uu}^+(y^+,y_l^+)
\,\mathrm{d}y_l^+ .
\label{eq3.4}
\end{equation}
\end{linenomath}
\par\noindent
This integration range lies entirely within the inverse-height scaling range in \eqref{eq3.3a} for $y^+\gtrsim100$ and $y/R\lesssim0.075$.
To interpret the wall-normal behaviour of this contribution, we consider the inverse-height scaling identified above.
Within the active portion,
$I_{uu}^+(y^+;y_l^+) \simeq \frac{A_a(y^*)}{y_l^+},$
where $A_a(y^*)$ is the $y^*$-dependent prefactor.
Using $y_l^+=y^+/y^*$ and $\mathrm{d}y_l^+/y_l^+=-\mathrm{d}y^*/y^*$ gives
\par
\begin{linenomath}
\begin{equation}
\langle uu\rangle_a^+
\simeq
\int_{0.15}^{0.5}
A_a(y^*)\,\mathrm{d}\ln y^*
\equiv C_a .
\label{eq3.5}
\end{equation}
\end{linenomath}
\par\noindent
Hence, the inverse-height scaling removes the dependence on the observation height $y$, leaving an integral solely over the fixed eddy-scaled interval $0.15\leq y^*\leq0.5$.
Accordingly, $C_a$ is independent of $y$ for a given Reynolds number, consistent with the approximately constant active contribution observed in figure~\ref{Fig07}($a$,$b$). 
In other words, as $y$ changes, both bounds of the contributing coherence-height range vary in proportion to $y$, so that the same relative portion of the self-similar hierarchy contributes at each observation location.

As shown in figure~\ref{Fig07}($a$,$b$), $\langle uu\rangle_a^+$ remains nearly constant over $100$--$150\lesssim y^+\lesssim0.1R^+$ for $Re_\tau=3000$ and $6000$.
The plateau thus extends slightly beyond $y/R\simeq0.075$, above which the upper part of the integration interval begins to exceed the inverse-height scaling range in \eqref{eq3.3a}.
Such a clear constant region is not observed at $Re_\tau=930$, consistent with the limited scale separation discussed above.
When expressed in inner units, the wall-normal extent of the plateau increases with Reynolds number, reflecting the broader range of coherence heights over which the inverse-height scaling is established.
The emergence of this constant cumulative contribution supports the inverse-height scaling identified within the active portion and demonstrates that the constant active contribution expected from the classical AEH can be recovered directly through the scale-resolved decomposition in $y_l$.

The approximately constant behaviour of $\langle uu\rangle_a^+$ is consistent with Townsend's description of active motions as a universal component characterized by the velocity and length scales $u_\tau$ and $y$, respectively \citep{Townsend61,Bradshaw67}.
A similar constant active contribution was reported by \citet{Deshpande21}.
As discussed in \S~\ref{sec:2.3}, however, their contribution was obtained as the residual remaining after removal of the estimated inactive contribution and does not explicitly resolve the wall-normal coherence height of the contributing motions.
In contrast, the present result is constructed by integrating the active portions of wall-coherent motions resolved according to $y_l$.
A corresponding scale-resolved analysis of wall-incoherent motions using an appropriate wall-normal coherence scale would help characterize the remaining active contributions but is beyond the scope of the present study.

It is also worth noting that the magnitude of $\langle uu\rangle_a^+$ exhibits a modest Reynolds-number dependence, with plateau values of approximately $1.4$ and $1.6$ at $Re_\tau=3000$ and $6000$, respectively.
The plateaus of $y_l^+I_{uu,\max}^+$ in the inset of figure~\ref{Fig05}($a$) give $A_a(y_a^*)\simeq1.16$ and $1.40$ at the corresponding Reynolds numbers.
As a simple estimate, approximating $A_a(y^*)$ by these peak values throughout the active interval gives $\langle uu\rangle_a^+\simeq1.40$ and $1.69$ from \eqref{eq3.5}, in reasonable agreement with the plateau values shown in figure~\ref{Fig07}($a$,$b$).
This agreement supports the interpretation that the nearly constant active contribution results from the inverse-height scaling of the individual wall-coherent motions.
However, the present range of $Re_\tau$ is insufficient to determine whether $A_a(y^*)$, or its average over the active interval, has a systematic Reynolds-number dependence.

Next, the cumulative inactive contribution within the identified self-similar range, $\langle uu\rangle_{ia}^+$, is evaluated by integrating $I_{uu}$ over $10y\leq y_l\leq0.25R$.
The lower bound follows from the inactive criterion $y^*=y/y_l\leq0.1$, whereas the upper bound is set by the outer limit of the $(y_l^+)^{-1}$ scaling identified in figures~\ref{Fig05}($c$,$d$) and \ref{Fig06}($b$).
Thus, the integration interval is consistent with the inactive portion defined in \eqref{eq3.3b}.
Within this range, the streamwise single-eddy intensity can be expressed as
$I_{uu}^+\simeq A_i(y^*)(y_l^+)^{-1}$, where $A_i(y^*)$ denotes the prefactor at each fixed $y^*$.
The cumulative inactive contribution is then given by
\par
\begin{linenomath}
\begin{equation}
\langle uu \rangle_{ia}^+(y^+)
\simeq
\int_{4y/R}^{0.1}
A_i(y^*)\,\mathrm{d}\ln y^* .
\label{eq3.6}
\end{equation}
\end{linenomath}
\par\noindent
This integration range lies within the inverse-height scaling range in \eqref{eq3.3b} for $y^+\gtrsim20$ and $y/R\lesssim0.025$.
If $A_i(y^*)$ varies only weakly over the corresponding inactive range, it may be approximated by a representative value $A_i$, yielding
\par
\begin{linenomath}
\begin{equation}
\langle uu \rangle_{ia}^+(y^+)
\simeq
A_i\ln\left(\frac{0.025R}{y}\right).
\label{eq3.7}
\end{equation}
\end{linenomath}
\par\noindent
Equation~\eqref{eq3.7} thus predicts a logarithmic wall-normal variation of the cumulative inactive contribution.
Figure~\ref{Fig07}($c$,$d$) shows $\langle uu\rangle_{ia}^+$ obtained directly from the extracted $I_{uu}$ (solid lines with circles).
A clear logarithmic variation is observed for both $Re_\tau=3000$ and $6000$, with good agreement between the two profiles.
The logarithmic slope is approximately $A_i\simeq0.65$, and the logarithmic behaviour extends roughly over $20 \lesssim y^+\lesssim0.025R^+$.
The nearly constant logarithmic slope is also consistent with a weak variation of $A_i(y^*)$ over the corresponding inactive range.
These results demonstrate that the cumulative contribution of the inactive footprints of wall-coherent motions can give rise to a logarithmic variation of the streamwise turbulence intensity, consistent with the classical AE description \citep{Townsend76,Perry82}; the magnitude of the logarithmic slope and its relation to the Townsend--Perry constant are discussed further in \S~\ref{sec:6.2}.

At a fixed $y^+$ within this range, \eqref{eq3.7} can be written as $\langle uu\rangle_{ia}^+ \sim \ln(Re_\tau/y^+)$, implying that $\langle uu\rangle_{ia}^+$ increases logarithmically with $Re_\tau$ \citep{Marusic97} and that the increment between two Reynolds numbers is approximately independent of $y^+$.
Consistent with this prediction, the difference between the $Re_\tau=3000$ and $6000$ profiles remains nearly constant over the logarithmic range but progressively decreases for $y^+\lesssim20$.
In particular, at $y^+\simeq15$, close to the near-wall peak of the total $\langle uu\rangle^+$, the increase is smaller than that observed within the logarithmic range. 
This location lies below the lower bound of the inactive inverse-height scaling range in \eqref{eq3.3b}, and the origin of this reduced increase is examined in \S~\ref{sec:4}.

Another important consequence of the identified scaling is that the logarithmic contribution from the inactive portions is confined to a relatively narrow near-wall region at the Reynolds numbers considered here.
Specifically, the upper bound $y^+\simeq0.025R^+$ corresponds to approximately $75$ and $150$ at $Re_\tau=3000$ and $6000$, respectively, consistent with figure~\ref{Fig07}($d$).
In the total streamwise turbulence intensity, this logarithmic variation can be obscured by contributions from smaller wall-coherent motions whose active portions occupy the same near-wall region.
This interpretation is consistent with the substantial contributions (i.e., active portions) from motions with $y_l^+\simeq100$--$200$, as shown in figure~\ref{Fig05}.
If the present upper limit of the inactive logarithmic range, $y^+\simeq0.025Re_\tau$, persists to higher Reynolds numbers, it would begin to overlap with the conventionally identified logarithmic region when it exceeds its lower bound.
Using $y^+\simeq3Re_\tau^{1/2}$ reported by \citet{Marusic13} gives $0.025Re_\tau\simeq3Re_\tau^{1/2}$, corresponding to $Re_\tau\simeq1.4\times10^4$.
This estimate suggests that the logarithmic variation associated with the inactive footprints of wall-coherent motions becomes discernible within the conventionally identified logarithmic region only at sufficiently high Reynolds numbers.

\subsection{Logarithmic variation of the cumulative contributions}
\label{sec:3.5}
We next consider the sum of the cumulative active and inactive contributions obtained in \S~\ref{sec:3.4}.
The coherence-height ranges used to define these contributions are $2y \leq y_l \leq y/0.15$ and $10y \leq y_l \leq 0.25R$, respectively, and therefore do not overlap.
The intervening range $y/0.15 < y_l < 10y$, corresponding to $0.1 < y^* < 0.15$, is excluded as the transition between the two portions.
As illustrated schematically in figure~\ref{Fig01}($d$), motions of different coherence heights can contribute simultaneously at a given observation location, with relatively shorter motions contributing through their active portions and taller motions through their inactive portions.
Where the constant active contribution and logarithmically varying inactive contribution are both established, their sum can be expressed using \eqref{eq3.5} and \eqref{eq3.7} as
\par
\begin{linenomath}
\begin{equation}
\begin{split}
\langle uu\rangle_a^+
+
\langle uu\rangle_{ia}^+
&\simeq
C_a
+
A_i\ln\left(\frac{0.025R}{y}\right) \\
&=
B_{ai}
-
A_i\ln\left(\frac{y}{R}\right),
\end{split}
\label{eq3.8}
\end{equation}
\end{linenomath}
\par\noindent
where $C_a$ denotes the approximately constant active contribution and $B_{ai}=C_a+A_i\ln(0.025)$.
Comparison with the classical AEH prediction in \eqref{eq1.3} shows that $A_i$ represents the contribution of the retained inactive wall-coherent motions to the logarithmic coefficient, whereas $B_{ai}$ is the corresponding intercept that includes the constant active contribution $C_a$.
Thus, the active contribution does not modify the logarithmic slope set by the inactive contribution, but provides an approximately constant additive offset.

At the Reynolds numbers considered here, however, the wall-normal ranges over which the constant active and logarithmic inactive contributions coexist remain limited.
The active plateau becomes established at $y^+\gtrsim100$--$150$, whereas the logarithmic variation of the inactive contribution extends only to $y^+\simeq0.025R^+=0.025Re_\tau$.
An overlap therefore requires $100$--$150\lesssim0.025Re_\tau$, corresponding approximately to $Re_\tau\gtrsim4000$--$6000$.
At $Re_\tau=6000$, this overlap is only beginning to emerge.
If the scaling ranges identified here persist at higher Reynolds numbers, the upper limit of the inactive logarithmic region will increase in inner units while the lower limit of the active plateau remains approximately fixed in inner units.
As a result, the wall-normal range over which the combined contribution follows \eqref{eq3.8} would broaden with increasing Reynolds number.

To examine the cumulative effect over a broader range of wall-coherent motions, we further consider $\langle uu\rangle_c^+$, obtained by integrating the single-eddy intensity over $2y\leq y_l\leq0.5R$, where the lower bound corresponds to $y^=0.5$ and the upper bound is set by the outer limit of the inverse-height scaling range.
Figure~\ref{Fig07}($c$,$d$) also shows $\langle uu\rangle_c^+$, while the insets present its logarithmic indicator function, $\Xi_c=-y^+\partial\langle uu\rangle_c^+/\partial y^+$.
An approximately constant indicator ($\Xi_c\simeq1.3$) is observed over $0.14\lesssim y/R\lesssim0.25$, indicating an outer logarithmic variation of $\langle uu\rangle_c^+$.
In this range, the integration corresponds to $0.28\lesssim y^*\leq0.5$, lying entirely within the active portions of relatively large wall-coherent motions.
This outer logarithmic variation is thus distinct from that in \eqref{eq3.8}, which arises from the logarithmically varying inactive contribution combined with a constant active offset.

This behaviour can be understood by applying the inverse-height scaling over the corresponding active range.
Since $2y\leq y_l\leq0.5R$ corresponds to $2y/R\leq y^*\leq0.5$, the cumulative contribution can be written as
\par
\begin{linenomath}
\begin{equation}
\langle uu\rangle_c^+
\simeq
\int_{2y/R}^{0.5}
A_a(y^*)\,\mathrm{d}\ln y^* .
\label{eq3.9}
\end{equation}
\end{linenomath}
\par\noindent
Accordingly, the logarithmic indicator becomes
\par
\begin{linenomath}
\begin{equation}
\Xi_c
=
-y^+\frac{\partial\langle uu\rangle_c^+}{\partial y^+}
\simeq
A_a(y^*=2y/R).
\label{eq3.10}
\end{equation}
\end{linenomath}
Thus, the approximately constant $\Xi_c$ indicates that $A_a(y^*)$ varies only weakly over $0.28\lesssim y^*\lesssim0.5$.
If $A_a(y^*)$ is approximated by a representative constant over this range, \eqref{eq3.9} reduces to $\langle uu\rangle_c^+ \simeq A_a\ln\left({0.25R}/{y}\right)$, consistent with the observed outer logarithmic variation.
The fixed outer bound $y_l\simeq0.5R$ is also consistent with figure~\ref{Fig05}($a$), where the inverse-height scaling begins to break down beyond this coherence height.
Hence, unlike the logarithmic variation arising from the accumulation of inactive footprints, the outer logarithmic behaviour here results from integrating the inverse-height-scaled active hierarchy up to its finite outer cutoff; this distinction is discussed further in \S~\ref{sec:6.2}.

Figure~\ref{Fig07}($d$) also exhibits an approximate plateau in $\Xi_c$ around $y^+=O(10^2)$, although it is less distinct than the outer plateau discussed above.
Given that the active contribution is nearly independent of $y$ over this range, the corresponding inner logarithmic tendency is attributed primarily to the accumulation of inactive footprints within the broader integration range $2y\leq y_l\leq0.5R$.
However, its logarithmic coefficient differs from that obtained from the isolated inactive contribution, particularly at $Re_\tau=6000$.
This difference likely reflects the inclusion of motions with $0.25R<y_l\leq0.5R$, whose lower inactive footprints lie beyond the inverse-height scaling range identified in figures~\ref{Fig05}($c$,$d$) and \ref{Fig06}($b$).
Consequently, the logarithmic variation produced by the self-similar inactive hierarchy can be modified by contributions from outer-scaled, non-self-similar wall-coherent motions, consistent with previous findings emphasizing the need to separate these contributions to reveal the underlying self-similar scaling \citep{Hwang20,Baars20}.

The inner and outer plateaus of $\Xi_c$ therefore arise from distinct parts of the wall-coherent hierarchy: the former is associated primarily with inactive footprints, whereas the latter results from the active portions of the largest motions bounded by the outer cutoff at $y_l\simeq0.5R$.
Under the present scaling, the inactive logarithmic range broadens in inner units with increasing Reynolds number, but its outer bound remains at approximately $y/R\simeq0.025$, well below the outer logarithmic range at $0.14\lesssim y/R\lesssim0.25$.
The appearance of these two separated wall-normal ranges in $\Xi_c$ is qualitatively reminiscent of the two invariant regions of the streamwise-velocity probability density function (p.d.f.) recently reported in high-Reynolds-number turbulent boundary layers \citep{Tsuji26} and in turbulent pipe and channel flows \citep{Tsuji26tsfp}.
These studies identified an inner invariant region over $100\lesssim y^+\lesssim200$, associated with a plateau-like variation of the total streamwise turbulence intensity, and an outer invariant region over $0.07\lesssim y/\delta\lesssim0.2$, where the total intensity exhibits the logarithmic variation expected from the AEH.
The two invariant regions become clearly distinguishable only at sufficiently high Reynolds numbers and are not separately resolved at lower Reynolds numbers.

The qualitative resemblance noted above suggests that the present decomposition may provide a scale-resolved interpretation of this two-region behaviour.
The absence of a clear logarithmic variation in the total turbulence intensity within the inner invariant region does not preclude the presence of an underlying logarithmic component.
The cumulative inactive footprints identified here exhibit such a variation over a partly overlapping inner-scaled range, although its signature in the total intensity could be weakened or offset by other $y$-dependent contributions, including wall-incoherent and outer-scaled motions.
Within the present decomposition, by contrast, the outer logarithmic tendency arises from the inverse-height-scaled active hierarchy bounded by its finite outer cutoff.
In this sense, the two invariant regions reported in \citet{Tsuji26} may partly reflect different statistical manifestations of the wall-coherent hierarchy, although establishing such a connection would require a comparable scale-resolved analysis at substantially higher Reynolds numbers.

\section{Near-wall streamwise turbulence intensity}
\label{sec:4}
The near-wall streamwise turbulence intensity attains a maximum near $y^+\simeq15$, and its peak magnitude, $\langle uu\rangle^+_{\max}$, increases with Reynolds number, although its asymptotic behaviour remains the subject of considerable debate.
As summarized by \citet{Hwangy24}, three principal Reynolds-number scalings have been proposed:
(i) unbounded logarithmic growth, $\langle uu\rangle^+_{\max}\sim\ln Re_\tau$, attributed to the increasing range of inactive AE contributions \citep{Marusic03,Marusic17};
(ii) convergence to a finite asymptote with a defect proportional to $1/U_\infty^+\sim1/\ln Re_\tau$, derived for zero-pressure-gradient boundary layers, where $U_\infty$ is the free-stream velocity \citep{Monkewitz15}; and
(iii) convergence to a finite asymptote through a defect power law, $\langle uu\rangle^+_{\max}=A-BRe_\tau^{-\beta}$.
For the third form, \citet{Chen21} proposed $\beta=1/4$ from a scaling argument based on the wall dissipation, whereas \citet{Pirozzoli24} obtained $\beta\simeq0.18$ empirically from the Reynolds-number dependence of the near-wall large-scale spectral contribution.
Despite their different asymptotic limits and physical foundations, all three forms can provide comparably good fits to the currently available data up to $Re_\tau=O(10^4)$, and thus data at substantially higher Reynolds numbers are required to distinguish their asymptotic trends \citep{Hwangy24,Jimenez24}.
The single-eddy decomposition developed in the preceding sections provides a complementary perspective on this problem by resolving the wall-coherent contribution to $\langle uu\rangle^+$ at $y^+=15$ according to $y_l$.
Hence, we examine the contributions from different ranges of $y_l$ and their Reynolds-number dependence at $y^+=15$.

\subsection{Coherence-height scaling of the near-wall contribution}
\label{sec:4.1}
Taking $y^+=15$ as the representative location of the near-wall peak, the streamwise turbulence intensity can be decomposed as
\par
\begin{linenomath}
\begin{equation}
\langle uu\rangle_{\max}^+
\equiv
\langle uu\rangle^+(y^+=15)
=
T_s^+ + T_{wc}^+ ,
\label{eq4.1}
\end{equation}
\end{linenomath}
\par\noindent
where $T_{wc}^+$ denotes the contribution from wall-coherent motions with coherence heights ranging from the near-wall reference location, $y_r^+=15$, to the outer scale:
\par
\begin{linenomath}
\begin{equation}
T_{wc}^+
=
\int_{y_{r}^+}^{R^+}
I_{uu}^+(15;y_l^+)
\,\mathrm{d}y_l^+.
\label{eq4.2}
\end{equation}
\end{linenomath}
\par\noindent
The remaining term $T_s^+(=\langle uu\rangle^+_{\max}-T_{wc}^+)$ denotes the residual contribution, including smaller-scale motions not represented by the present decomposition.
Since $y^+=15$ lies very close to the wall, this residual is expected to contain substantial contributions from viscous-scale motions.

\begin{figure}
\centerline{\includegraphics[trim=2.5cm 1.0cm 2.5cm 0.2cm,width=14cm]{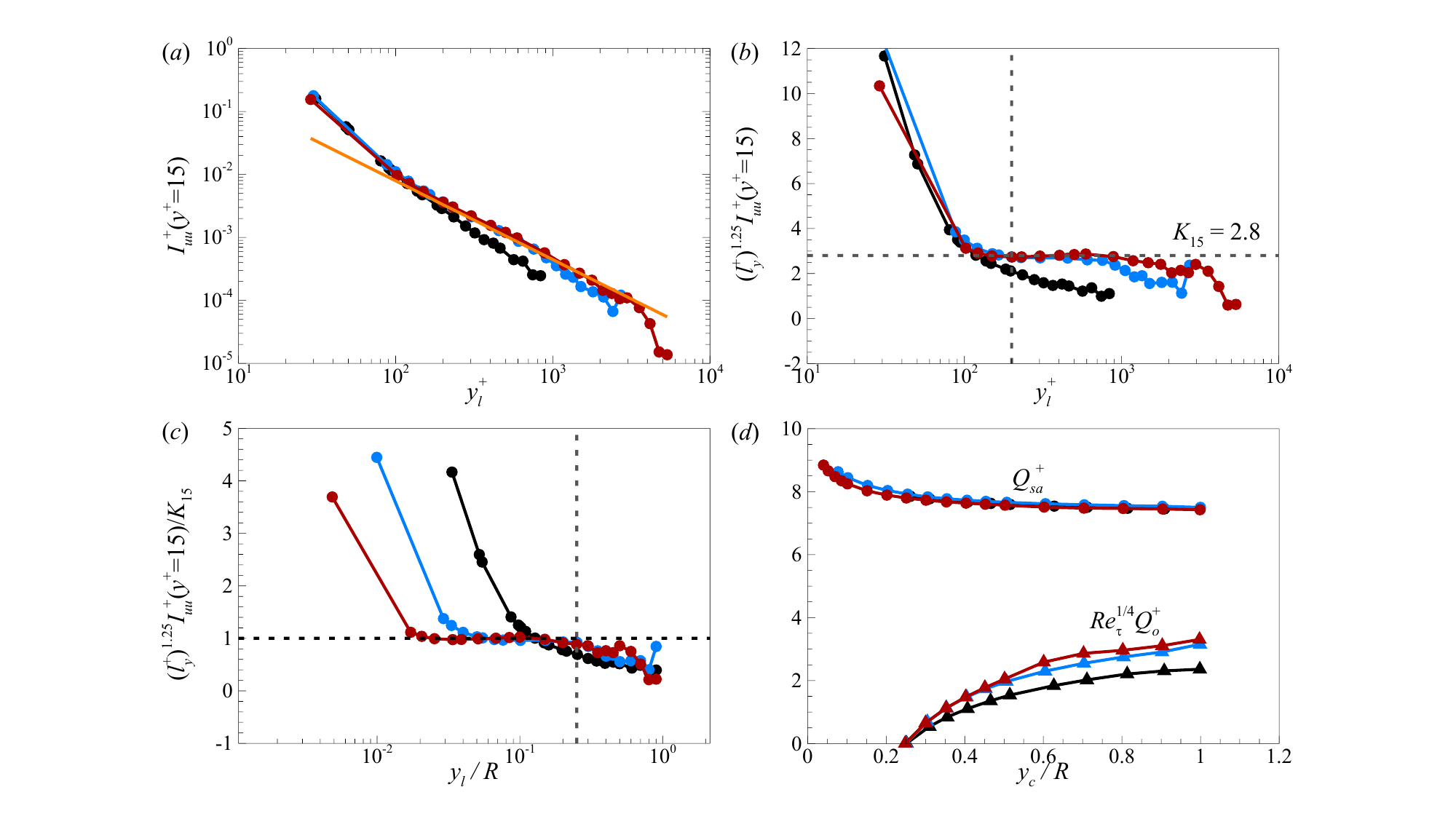}}
 \caption{Single-eddy contributions to the streamwise turbulence intensity at $y^+=15$.
($a$) $I_{uu}^+(15;y_l^+)$ as a function of the wall-normal coherence height $y_l$.
The solid orange line indicates the scaling $I_{uu}^+\sim(y_l^+)^{-1.25}$.
($b$) Premultiplied intensity, $(y_l^+)^{1+\beta}I_{uu}^+(15;y_l^+)$, plotted against $y_l^+$, where $\beta=1/4$.
The vertical dashed line denotes $y_l^+=200$, marking the onset of the $(y_l^+)^{-1.25}$ scaling, and the horizontal dashed line denotes the plateau level $K_{15}\simeq2.8$.
($c$) The premultiplied intensity normalized by $K_{15}$, plotted against $y_l/R$.
The vertical dashed line denotes $y_l/R=0.25$, above which the outer-cutoff behaviour becomes apparent.
($d$) Cumulative contributions obtained by varying the upper coherence-height limit $y_c/R$: $Q_{sa}^+$ (circles) and the compensated outer contribution $Re_\tau^{1/4}Q_o^+$ (triangles).
At $y_c/R=1$, these quantities correspond to $T_s^+ + T_a^+$ and $Re_\tau^{1/4}T_o^+$, respectively.
The colour coding is as specified in table~\ref{tab:Table1}.}
 \label{Fig08}
\end{figure}

To examine the coherence-height dependence of the integrand in \eqref{eq4.2}, figure~\ref{Fig08}($a$) shows $I_{uu}^+(15;y_l^+)$ as a function of $y_l^+$.
The profiles show good agreement across Reynolds numbers, particularly for $y_l^+\lesssim200$, where the contribution is dominated by relatively small wall-coherent motions.
For $y_l^+\gtrsim200$, however, $I_{uu}^+(15;y_l^+)$ decreases approximately as $(y_l^+)^{-1.25}$ over an intermediate range, as indicated by the solid orange line.
This decay is slightly steeper than the inverse-height scaling, $I_{uu}^+\sim(y_l^+)^{-1}$, identified in figure~\ref{Fig05} and \ref{Fig06}.
Since a fixed $y^+=15$ samples progressively smaller $y^*=y/y_l$ as $y_l$ increases, the steeper decay reflects an additional attenuation of the near-wall inactive footprint with increasing coherence height.
As $y_l$ approaches $O(R)$, $I_{uu}^+$ departs from this power-law behaviour (solid orange line) and decreases more rapidly, indicating the onset of an outer-scale cutoff.
The cutoff behaviour is similar for $Re_\tau=3000$ and $6000$, whereas the departure occurs at a somewhat smaller $y_l/R$ for $Re_\tau=930$.

Unlike the inverse-height scaling identified at fixed $y^*$, the present scaling is evaluated at the fixed near-wall location $y^+=15$, thereby sampling the lower inactive portion of progressively larger wall-coherent motions.
To account for the resulting additional attenuation, we introduce an exponent $\beta$ and write
\par
\begin{linenomath}
\begin{equation}
I_{uu}^+(15;y_l^+)
\simeq
K_{15}(y_l^+)^{-(1+\beta)}
G(y_l/R),
\label{eq4.3}
\end{equation}
\end{linenomath}
\par\noindent
where $K_{15}$ is the coefficient of the power-law contribution, $\beta$ represents the additional attenuation of the inactive footprint relative to the inverse-height scaling, and $G(y_l/R)$ represents the outer-scale cutoff.
The present data give $\beta\simeq1/4$, yielding $1+\beta\simeq1.25$.
The cutoff function is approximately unity over the power-law range and decreases as $y_l$ approaches $O(R)$, accounting for the outer-scale departure from the self-similar hierarchy discussed in connection with figure~\ref{Fig05}.
The physical origin of the near-wall exponent $\beta$ is discussed further in \S\ref{sec:4.3}.

To test the representation in \eqref{eq4.3}, figure~\ref{Fig08}($b$) shows the premultiplied intensity, $(y_l^+)^{1.25}I_{uu}^+(15;y_l^+)$, as a function of $y_l^+$.
The data at $Re_\tau=3000$ and $6000$ exhibit an approximately constant plateau over $200\lesssim y_l^+\lesssim0.25R^+$, with a value of $K_{15}\simeq2.8$, as indicated by the horizontal dashed line.
In contrast, no clear plateau is observed at $Re_\tau=930$, where the intensity begins to decrease shortly beyond $y_l^+\simeq200$.
To examine the outer-scale departure from this plateau, figure~\ref{Fig08}($c$) shows the premultiplied intensity normalized by $K_{15}$ as a function of $y_l/R$.
The normalized intensity remains approximately unity up to $y_l/R\approx0.25$ and subsequently decreases, with the decreasing branches exhibiting reasonable agreement across Reynolds numbers.
Thus, the data support the representation in \eqref{eq4.3}, with $\beta\simeq1/4$ and $G$ characterized primarily by the outer-scaled coherence height $y_l/R$.
The absence of a clear plateau at $Re_\tau=930$ is consistent with insufficient scale separation, since $0.25R^+\approx230$, leaving essentially no interval between the lower limit $y_l^+\simeq200$ and the onset of the outer-scale cutoff.

Based on the scaling behaviour identified above, the wall-coherent contribution $T_{wc}^+$ can be decomposed as
\par
\begin{linenomath}
\begin{equation}
T_{wc}^+
=
T_a^+ + T_i^+ + T_o^+ ,
\label{eq4.4}
\end{equation}
\end{linenomath}
\par\noindent
where $T_a^+$, $T_i^+$, and $T_o^+$ denote the lower-coherence-height contribution dominated by the active portion, the near-wall inactive-footprint contribution, and the outer-scale contribution, respectively.
Here, $T_i^+$ is associated with $200\lesssim y_l^+\lesssim c_oR^+$, with $c_o\simeq0.25$, whereas $T_o^+$ corresponds to $c_oR^+\lesssim y_l^+\lesssim R^+$.
Integrating \eqref{eq4.3} over these ranges gives
\par
\begin{linenomath}
\begin{subequations}
\label{eq4.5}
\begin{align}
T_i^+
&=
T_{i,\infty}^+-D_iRe_\tau^{-\beta},
\label{eq4.5a}\\
T_o^+
&=
D_oRe_\tau^{-\beta}.
\label{eq4.5b}
\end{align}
\end{subequations}
\end{linenomath}
\par\noindent
where $T_{i,\infty}^+$ denotes the finite asymptotic limit of the inactive-footprint contribution as $Re_\tau\to\infty$, while $D_i$ and $D_o$ are determined by $K_{15}$, $c_o$ and the outer-cutoff function $G$ (see Appendix~\ref{app:peak_scaling}).

Given the approximate Reynolds-number invariance of $K_{15}$ and $G$ observed above, and assuming that $T_s^++T_a^+$ approaches a Reynolds-number-independent limit, as expected for the small-scale and predominantly active contributions, the near-wall intensity takes the form
\par
\begin{linenomath}
\begin{equation}
\langle uu\rangle_{\max}^+
=
A-BRe_\tau^{-\beta},
\label{eq4.6}
\end{equation}
\end{linenomath}
\par\noindent
where $A=(T_s^++T_a^+)_{\infty}+T_{i,\infty}^+$ is the asymptotic near-wall intensity and $B=D_i-D_o$ is the corresponding defect coefficient.
Thus, the observed $\beta\simeq1/4$ leads to a finite asymptote with a $Re_\tau^{-1/4}$ defect.

To examine the Reynolds-number dependence underlying \eqref{eq4.6}, figure~\ref{Fig08}($d$) shows the combined residual and lower-coherence-height contribution, $Q_{sa}^+$ (circles), and the compensated outer contribution, $Re_\tau^{1/4}Q_o^+$ (triangles), as the upper integration limit $y_c/R$ is progressively increased.
Here, $Q_{sa}^+$ is obtained by subtracting the wall-coherent contribution over $200\leq y_l^+\leq y_c^+$ from $\langle uu\rangle_{\max}^+$, while $Q_o^+$ denotes the contribution integrated over $0.25R^+\leq y_l^+\leq y_c^+$.
As $y_c/R\to1$, these quantities approach $T_s^++T_a^+$ and $T_o^+$, respectively.

As seen, $Q_{sa}^+$ approaches a similar level across the Reynolds numbers, with $T_s^++T_a^+\simeq7.45$, supporting the assumption that this combined contribution becomes approximately Reynolds-number independent.
The outer contribution is much smaller ($T_o^+\lesssim0.4$), and decreases with increasing Reynolds number (not shown here).
When premultiplied by $Re_\tau^{1/4}$, the curves (triangles) at $Re_\tau=3000$ and $6000$, despite some differences around $y_c/R\simeq0.6$--$0.7$, approach a comparable limiting value ($D_o\simeq3.0$) as $y_c/R\to1$.
This behaviour is consistent with the predicted scaling in \eqref{eq4.5b}. 
At $Re_\tau=930$, such convergence is less evident, consistent with the insufficient scale separation discussed above.
Together with the approximate Reynolds-number invariance of $T_s^++T_a^+$, these results support the finite-limit defect form in \eqref{eq4.6}, although the scaling of $T_o^+$ requires confirmation over a wider Reynolds-number range.

\subsection{Comparison with existing near-wall scaling models}
\label{sec:4.2}
The asymptotic form obtained here is consistent with the saturation of the near-wall intensity proposed by \citet{Chen21} and \citet{Pirozzoli24}.
In particular, the present value $\beta\simeq1/4$ coincides with the defect exponent proposed by \citet{Chen21}, whereas \citet{Pirozzoli24} obtained a somewhat smaller value, $\beta\simeq0.18$, from the scaling of the energy spectra.
Despite the identical defect exponent in the former case, however, the physical basis of the present result is different.
\citet{Chen21} derived the $Re_\tau^{-1/4}$ defect by assuming that the wall dissipation approaches a limiting value of $1/4$ and that its finite-Reynolds-number defect is governed by an outer Kolmogorov length scale associated with near-wall bursting.
In the present analysis, by contrast, the defect exponent emerges from the coherence-height scaling of individual wall-coherent contributions, while the coefficients $A$ and $B$ are explicitly related to the active, inactive-footprint and outer-scale contributions.
Thus, the same $Re_\tau^{-1/4}$ defect form is recovered without prescribing an asymptotic value of the wall dissipation.

A quantitative comparison can also be made for the coefficients of the defect law.
Using the observed values $K_{15}\simeq2.8$, $T_s^++T_a^+\simeq7.45$, and $D_o\simeq3.0$, the present decomposition gives $A\simeq10.4$ and $B\simeq12.8$, compared with $A_{\mathrm{CS}}\simeq11.5$ and $B_{\mathrm{CS}}\simeq19.3$ from \citet{Chen21}.
Thus, the predicted asymptotic level is reasonably close, whereas a larger quantitative difference remains in the defect coefficient.
It should be noted, however, that the present estimates assume that the observed plateau value remains Reynolds-number independent at $K_{15}\simeq2.8$.
As seen in figure~\ref{Fig08}($c$), the premultiplied intensity at $Re_\tau=930$ remains slightly below those at the two higher Reynolds numbers.
Although this difference may reflect the limited scale separation and the early influence of the outer-scale cutoff, it leaves open the possibility that some Reynolds-number dependence of $K_{15}$ persists.
Such a dependence does not alter the leading $Re_\tau^{-1/4}$ defect provided that $K_{15,\infty}-K_{15}=O(Re_\tau^{-1/4})$ or faster; an $O(Re_\tau^{-1/4})$ correction modifies the defect coefficient $B$, whereas a slower approach would alter the leading Reynolds-number dependence.
Since $K_{15}$ enters both the asymptotic inactive contribution and the defect coefficient, a further increase towards its asymptotic value would increase both $A$ and $B$, thereby reducing their differences from the values of \citet{Chen21}.
For example, if $K_{15,\infty}\simeq3.8$ and the cutoff function $G$ remains unchanged, the present model would give $A\simeq11.5$ and $B\simeq17.4$.
The asymptotic value and Reynolds-number dependence of $K_{15}$ cannot, however, be established from the presently available Reynolds-number range and require further investigation at substantially higher Reynolds numbers.

Although the $Re_\tau^{-1/4}$ model of \citet{Chen21} describes the available data well, the physical assumptions underlying its derivation remain uncertain as $Re_\tau\to\infty$.
As discussed by \citet{Hwangy24}, there is no compelling physical reason for the wall dissipation to be bounded by the asymptotic maximum of the inner-scaled turbulence production, $1/4$, since near-wall dissipation need not be determined solely by local production.
Motions originating farther from the wall can transport energy towards the near-wall region and contribute to its dissipation.
This possibility is already implicit in the classical active--inactive framework: \citet{Bradshaw67} argued that inactive portions of large-scale motions can enhance dissipation in the viscous sublayer through turbulent-energy diffusion.
In addition, there is no clear evidence that near-wall bursting transports turbulence to the outer region as $Re_\tau\to\infty$.
As pointed out by \citet{Hwangy24}, the wall-normal transport distance associated with the outer Kolmogorov timescale becomes increasingly small relative to the outer length scale as Reynolds number increases.
Hence, the assumed transport of near-wall turbulence across an outer-scaled distance becomes increasingly difficult to sustain as $Re_\tau\rightarrow\infty$.

\subsection{Physical origin of the $\beta=1/4$ scaling}
\label{sec:4.3}
A possible physical interpretation of the measured $\beta\simeq1/4$ can be obtained in terms of the eddy-height dependence of the characteristic dissipation scale.
The classical active--inactive picture suggests that the near-wall energy associated with large wall-attached motions need not be generated locally.
In particular, \citet{Bradshaw67} argued that inactive large-scale motions contribute little directly to the local Reynolds shear stress and turbulence production, whereas their enhanced dissipation in the viscous sublayer is supplied by turbulent-energy diffusion from the outer active region.
Similarly, \citet{Nikora99}, based on Townsend's equilibrium-layer framework \citep{Townsend61}, described wall turbulence as a superposition of energy cascades initiated at different wall-normal locations, with the characteristic size of each energy-containing motion proportional to its distance from the wall.
Here, we apply a similar argument to the hierarchy of wall-coherent motions characterized by $y_l$.
Similar to the logarithmic-interval representation in \eqref{eq3.1} and \eqref{eq3.2}, we define the contribution at $y^+=15$ per unit logarithmic interval of coherence height as
\par
\begin{linenomath}
\begin{equation}
J_{uu}^+(y_l^+)
\equiv
y_l^+ I_{uu}^+(15;y_l^+).
\label{eq4.7}
\end{equation}
\end{linenomath}
\par\noindent
In the ideal AE hierarchy, $I_{11}^{AE}$ approaches a constant within the inactive region, while the inverse-height weighting $W(y_l)\sim y_l^{-1}$ gives $W(y_l)I_{11}^{AE}\sim y_l^{-1}$, as expressed in \eqref{eq1.4}.
Since the present $I_{uu}^+$ corresponds to this weighted integrand, the ideal-AE counterpart of \eqref{eq4.7} is $J_{uu,\mathrm{AE}}^+\approx\mathrm{const.}$

Taking $y_l$ as the characteristic wall-normal size of a wall-coherent motion, Townsend's equilibrium-layer scaling suggests a characteristic dissipation rate $\varepsilon_l\sim u_\tau^3/y_l$.
The corresponding Kolmogorov length scale is
\par
\begin{linenomath}
\begin{equation}
\eta_l^+
=
\frac{u_\tau}{\nu}
\left(\frac{\nu^3}{\varepsilon_l}\right)^{1/4}
\sim
(y_l^+)^{1/4}.
\label{eq4.8}
\end{equation}
\end{linenomath}
\par\noindent
Motivated by Bradshaw's picture in which near-wall inactive energy is supplied non-locally by larger energetic motions and ultimately encounters viscous dissipation near the wall, we hypothesize that the strength of the near-wall inactive footprint depends on the dissipative scale associated with a wall-coherent motion of coherence height $y_l$. 
Specifically, the footprint is expected to weaken progressively as the corresponding Kolmogorov scale, $\eta_l$, increases relative to the viscous length. 
As the simplest representation, we assume that the leading attenuation is proportional to the inverse of this scale ratio, giving
\par
\begin{linenomath}
\begin{equation}
F_\nu(y_l^+)
\sim
\frac{\delta_\nu}{\eta_l}
=
\frac{1}{\eta_l^+}
\sim
(y_l^+)^{-1/4}.
\label{eq4.9}
\end{equation}
\end{linenomath}
\par\noindent
Applying this attenuation to the ideal AE contribution per unit logarithmic interval gives
\par
\begin{linenomath}
\begin{equation}
J_{uu}^+
\sim
J_{uu,\mathrm{AE}}^+ F_\nu
\sim
(y_l^+)^{-1/4},
\label{eq4.10}
\end{equation}
\end{linenomath}
\par\noindent
and hence, using \eqref{eq4.7},
\par
\begin{linenomath}
\begin{equation}
I_{uu}^+(15;y_l^+)
=
\frac{J_{uu}^+}{y_l^+}
\sim
(y_l^+)^{-1}(y_l^+)^{-1/4}
=
(y_l^+)^{-5/4}.
\label{eq4.11}
\end{equation}
\end{linenomath}
\par\noindent
Thus, the exponent unity represents the inverse-height scaling of the ideal AE hierarchy, whereas the additional $\beta=1/4$ may be interpreted as a $y_l$-dependent viscous attenuation of the near-wall inactive footprint associated with $\eta_l$.

It is useful to distinguish the present interpretation from the argument of \citet{Chen21}.
Formally, the Kolmogorov scale in \eqref{eq4.8} is analogous to the outer Kolmogorov scale employed by \citet{Chen21}, with the outer length $R$ replaced by the coherence height $y_l$.
Here, however, a distinct dissipative scale is associated with each wall-coherent motion rather than invoking a single Kolmogorov scale based on the outer length $R$.
This eddy-specific application of the equilibrium-layer scaling is consistent with Townsend's framework and the interpretation of \citet{Nikora99}, in which energy-containing motions at different wall-normal locations initiate corresponding energy cascades.
Consequently, unlike the model of \citet{Chen21} discussed in \S~\ref{sec:4.2}, the present argument requires neither a wall dissipation bounded by the maximum local production nor the transport of near-wall turbulence over an outer-scaled wall-normal distance.
Instead, it considers wall-attached motions spanning different coherence heights and describes the attenuation of their near-wall inactive footprints at a fixed near-wall location.
This scale-dependent attenuation is also consistent with the observation that viscous effects progressively limit the near-wall influence of large energy-containing motions \citep{Hwangy24}.

Nevertheless, the specific form of the attenuation factor in \eqref{eq4.9} remains phenomenological.
Although the scaling $\eta_l^+\sim(y_l^+)^{1/4}$ follows directly from the equilibrium-layer estimate $\varepsilon_l\sim u_\tau^3/y_l$, the dependence of the near-wall inactive footprint on $\delta_\nu/\eta_l$ is not derived from a governing energy balance.
\citet{Bradshaw67} related the near-wall dissipation of inactive energy to turbulent-energy diffusion, whereas \citet{Yang26} formulated an AEH-based inactive-energy budget involving mean convection, viscous diffusion, dissipation, and energy transfer from active motions.
Such an energy-budget framework may provide a basis for examining the eddy-height-dependent attenuation proposed here.
More generally, the attenuation can be expressed as $F_\nu\sim(\delta_\nu/\eta_l)^p$, yielding an additional coherence-height exponent $\beta=p/4$.
The present result, $\beta\simeq1/4$, therefore corresponds to $p\simeq1$, indicating an approximately linear dependence on $\delta_\nu/\eta_l$ over the coherence-height range in which the $(y_l^+)^{-5/4}$ scaling is observed.
For comparison, the exponent $\beta\simeq0.18$ reported by \citet{Pirozzoli24} and employed in the spectral model of \citet{Pirozzoli26} would correspond to $p\simeq0.72$ within the present interpretation, representing a somewhat weaker dependence on the dissipative-scale ratio.
The exponent of \citet{Pirozzoli24}, however, originates from the overlap scaling of the near-wall azimuthal-wavenumber spectra and the associated Reynolds-number defect, rather than from the coherence-height dependence considered here.
A corresponding spectral analysis of the present single-eddy contributions will be presented in Part~2.
Whether the approximately linear dependence identified here persists at asymptotically high Reynolds numbers or can be derived from the scale-dependent energy budget of individual wall-attached motions remains to be established.

\section{Outer secondary peak}
\label{sec:5}
We now consider whether the single-eddy framework can account for the emergence of an outer secondary peak in $\langle uu\rangle^+$.
Evidence for such a peak has been reported in several high-Reynolds-number pipe-flow experiments, although its prominence and Reynolds-number dependence vary among datasets and facilities.
Using the Princeton SuperPipe data, \citet{Vallikivi15} showed that the inner-scaled wall-normal location of the outer spectral peak, which closely follows that of the outer peak in $\langle uu\rangle^+$, scales approximately as $Re_\tau^{1/2}$ at lower Reynolds numbers but approaches $y^+\simeq400$ for $Re_\tau\gtrsim2\times10^4$.
In the CICLoPE facility, the PIV measurements of \citet{Willert17} showed a distinct outer peak for $Re_\tau\gtrsim2\times10^4$, whereas the X-wire measurements of \citet{Orlu17} exhibited only a weak secondary peak that became approximately plateau-like after spatial-resolution correction.
Similarly, the Hi-Reff measurements of \citet{Ono22}, extending to $Re_\tau\simeq20\,750$, showed a progressive Reynolds-number increase in $\langle uu\rangle^+$ within $y^+\simeq200$--$400$, without a clearly distinguishable secondary peak.
These observations suggest that the Reynolds-number increase in the outer-region intensity is a robust feature, whereas the appearance and prominence of a distinct secondary maximum remain sensitive to measurement resolution and methodology.

Recent DNS-based analyses have explored how this outer-region feature may evolve beyond the Reynolds numbers currently accessible to DNS.
\citet{Jimenez24} argued that large-scale fluctuations become increasingly important with increasing Reynolds number and extend over a wall-normal distance of $O(R/5)$, eventually producing a secondary maximum away from the wall.
More recently, \citet{Pirozzoli26} attributed the emergence of the outer peak to the increasing relative contribution of intermediate-scale wall-attached motions, whereas the contribution from the largest outer-scaled motions decreases with Reynolds number at fixed $y^+$.
Their extrapolation predicts that a clearly distinguishable outer peak develops only at sufficiently high Reynolds numbers.
Motivated by these results, we use the present single-eddy decomposition to isolate the contributions from different coherence-height ranges and assess their possible roles in the emergence of the outer secondary peak.

To examine the contribution of the wall-coherent motions to the outer-region intensity, we first consider their cumulative contribution over the range $y_l^+\geq200$, defined as
\par
\begin{linenomath}
\begin{equation}
\langle uu\rangle_{wc,200}^+(y)
=
\int_{200}^{R^+}
I_{uu}^+(y;y_l^+)
\,\mathrm{d}y_l^+ .
\label{eq5.1}
\end{equation}
\end{linenomath}
\par
\begin{figure}
\centerline{\includegraphics[trim=2.5cm 1.0cm 2.5cm 0.2cm,width=14cm]{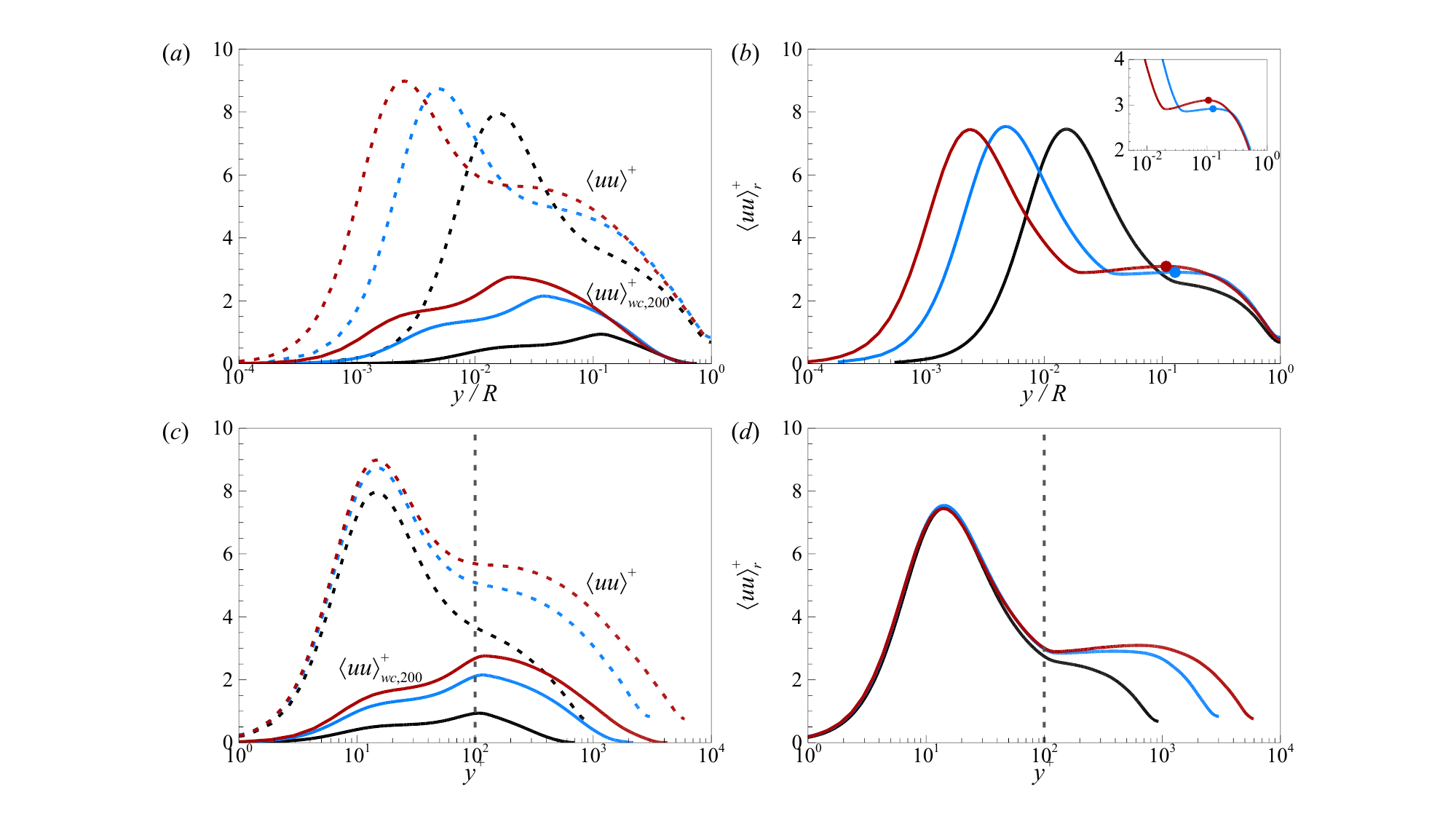}}
 \caption{($a$,$c$) Wall-normal profiles of the total streamwise turbulence intensity, $\langle uu\rangle^+$ (dashed lines), and the integrated wall-coherent contribution, $\langle uu\rangle_{wc,200}^+$ (solid lines), plotted against ($a$) $y/R$ and ($c$) $y^+$.
($b$,$d$) Wall-normal profiles of the residual intensity, $\langle uu\rangle_r^+=\langle uu\rangle^+-\langle uu\rangle_{wc,200}^+$, plotted against ($b$) $y/R$ and ($d$) $y^+$.
The inserted circles denote the outer-peak-like bulge observed in $\langle uu\rangle_{r}^+$ at $Re_\tau=3000$ and $6000$, which may contribute to the emergence of the outer peak in the total intensity.
A magnified view of this region is shown in the inset of ($b$).
In ($c$,$d$), the vertical dashed line denotes $y^+=100$.
The colour coding is as specified in table~\ref{tab:Table1}.}
 \label{Fig09}
\end{figure}

\noindent
At $y^+=15$, this contribution corresponds to $T_i^++T_o^+$ in the decomposition introduced in \S~\ref{sec:4}.
Figures~\ref{Fig09}($a$,$c$) show the wall-normal profiles of $\langle uu\rangle_{wc,200}^+$ (solid lines) plotted against ($a$) $y/R$ and ($c$) $y^+$, together with the total streamwise turbulence intensity $\langle uu\rangle^+$ (dashed lines) for comparison.
As seen in figure~\ref{Fig09}($c$), $\langle uu\rangle_{wc,200}^+$ increases with Reynolds number in inner scaling and extends progressively towards larger $y^+$.
This behaviour reflects the increasing range of coherence heights included in \eqref{eq5.1} as $R^+$ increases.
In outer scaling, however, the profiles exhibit reasonable collapse beyond $y/R\simeq0.2$--$0.3$, as shown in figure~\ref{Fig09}($a$).
This region is predominantly influenced by wall-coherent motions with $y_l=O(R)$, consistent with the outer-scaled behaviour of the largest wall-coherent motions identified in \S~\ref{sec:3}.
In addition, $\langle uu\rangle_{wc,200}^+$ exhibits a maximum near $y^+\simeq100$, whose location varies only weakly with Reynolds number.
This maximum is associated with the lower coherence-height limit ($y_l^+=200$) imposed in \eqref{eq5.1}.
As shown in \S~\ref{sec:3.3}, the active region of an individual wall-coherent motion extends to $y^*\simeq0.5$.
Thus, for the smallest motion included in \eqref{eq5.1}, the upper extent of the active region corresponds to $y^+\simeq100$, providing a natural inner-scaled location around which the cumulative contribution turns over.

Figures~\ref{Fig09}($b$,$d$) show the residual intensity, $\langle uu\rangle_r^+$ ($=\langle uu\rangle^+-\langle uu\rangle_{wc,200}^+$), plotted against ($b$) $y/R$ and ($d$) $y^+$.
The residual represents the contribution not accounted for by wall-coherent motions with $y_l^+\geq200$ and therefore includes both smaller wall-coherent motions with $y_l^+<200$ and wall-incoherent motions over a broad range of scales.
At $y^+=15$, $\langle uu\rangle_{wc,200}^+=T_i^++T_o^+$ in \S~\ref{sec:4}, and hence $\langle uu\rangle_r^+=T_s^++T_a^+$.
Since the wall-incoherent contribution is expected to be small near the wall, the residual in the near-wall region (e.g., $y^+\lesssim100$) is expected to contain a substantial contribution from the active portions of smaller wall-coherent motions with $y_l^+<200$, whereas wall-incoherent contributions become increasingly important with increasing distance from the wall.
As shown in figure~\ref{Fig09}($d$), the residual profiles collapse remarkably well for $y^+\lesssim100$, indicating that removing the contribution from wall-coherent motions with $y_l^+\geq200$ recovers an approximately universal near-wall intensity over this range.

Beyond $y^+\simeq100$, the residual profiles vary only weakly over an extended wall-normal range, with $\langle uu\rangle_r^+\simeq2.8$--$3.0$.
As discussed in \S~\ref{sec:2.3}, the residual is conceptually similar to that examined by \citet{Deshpande21}.
Their integrated residual has a comparable magnitude of approximately $2.7$ at the wall-normal locations considered, although it was obtained using a different decomposition method and exhibits some variation with wall-normal position.
Importantly, wall-incoherent motions need not be inactive in the Townsend sense, since active motions are defined by their contribution to turbulent momentum transfer at the observation location \citep{Townsend61}, rather than by their coherence with a near-wall reference.
In this sense, a wall-incoherent motion localized around a given wall-normal position can provide a locally active contribution despite being incoherent with the near-wall reference.
As the observation location moves away from the wall, contributions from such locally active wall-incoherent motions may provide one explanation for the relatively weak wall-normal variation of $\langle uu\rangle_r^+$.

A closer inspection, however, reveals a weak outer-region feature, as highlighted in the inset of figure~\ref{Fig09}($b$).
At $Re_\tau=3000$, it appears approximately plateau-like, whereas at $Re_\tau=6000$ a weak hump becomes discernible; the corresponding locations are indicated by the circles.
These observations may be related to the emergence of the outer peak discussed below.
Farther from the wall, $\langle uu\rangle_r^+$ decreases rapidly beyond $y/R\simeq0.2$--$0.3$.

A comparison of the two contributions, $\langle uu\rangle_{wc,200}^+$ and $\langle uu\rangle_r^+$, further reveals distinct Reynolds-number dependences.
For $y^+\lesssim100$, the increase of $\langle uu\rangle_{wc,200}^+$ with Reynolds number is relatively modest, while $\langle uu\rangle_r^+$ remains nearly Reynolds-number independent.
The weak Reynolds-number dependence of both contributions over this range is consistent with the tendency towards saturation of the near-wall peak discussed in \S~\ref{sec:4.2}.
At larger wall distances, however, the Reynolds-number growth of the total intensity is predominantly associated with wall-coherent motions with $y_l^+\geq200$, whereas $\langle uu\rangle_r^+$ changes comparatively little in magnitude.

The maximum of $\langle uu\rangle_{wc,200}^+$ near $y^+\simeq100$ contributes to the gradual flattening of the total intensity profile, producing shoulder-like behaviour by $Re_\tau\simeq6000$.
As Reynolds number increases, however, this contribution grows primarily through its extension towards larger $y^+$, rather than through a localized accumulation of energy around its maximum.
Indeed, the Reynolds-number increment in $\langle uu\rangle_{wc,200}^+$ between $Re_\tau=3000$ and $6000$ is largest at $y^+\simeq835$, well beyond the maximum near $y^+\simeq100$ and within the decaying outer portion of the wall-coherent contribution.
As a result, within the present Reynolds-number range, the wall-coherent contribution promotes the formation of an outer shoulder but is insufficient by itself to produce a distinct outer secondary maximum in the total intensity profile.

From this perspective, the weak outer-region hump in $\langle uu\rangle_r^+$ would be relevant to the eventual emergence of an outer secondary peak.
Its location, indicated by the circles in figure~\ref{Fig09}($b$), shifts slightly towards the wall in outer coordinates, from $y/R\simeq0.13$ at $Re_\tau=3000$ to $y/R\simeq0.11$ at $Re_\tau=6000$.
The residual profile is nearly plateau-like at $Re_\tau=3000$ but becomes more convex around the indicated location at $Re_\tau=6000$, suggesting the gradual development of a localized outer-region feature.
If this tendency persists at higher Reynolds numbers, its superposition with the broadly distributed wall-coherent contribution could contribute to the formation of a distinct outer secondary maximum.
The shift of this feature towards the wall in outer coordinates is qualitatively similar to that reported for the outer spectral and turbulence-intensity peaks in high-Reynolds-number pipe flow.
However, as noted above, \citet{Vallikivi15} showed that the peak location eventually approaches an approximately fixed position in inner units.
The present Reynolds-number range is insufficient to infer an asymptotic scaling for the location of the residual feature.

An interesting comparison can be made with the spectral analysis of \citet{Pirozzoli26}, who attributed the emergence of the outer peak to the increasing contribution of intermediate-scale motions associated with the attached-eddy spectral range.
In their analysis, this range is identified from a self-similar spectral organisation characterised by a linear relation between wall-normal distance and wavelength, $\lambda_\theta\propto y$, and interpreted in the context of the AEH.
This spectrum-based classification is not equivalent to the present coherence-based decomposition, in which wall-coherent motions are explicitly identified through their coherence with the near-wall reference location at $y^+=15$.
Consequently, motions within the intermediate spectral range identified by \citet{Pirozzoli26} need not be wall-coherent according to the present criterion.
If relatively large-scale motions retain the self-similar relation between wall-normal position and wavelength while remaining incoherent with the near-wall reference, their contributions would remain in $\langle uu\rangle_r^+$ rather than in $\langle uu\rangle_{wc,200}^+$.
Hence, such motions could contribute to the localized outer-region feature observed in the residual and eventually to the emergence of a distinct outer secondary peak.
This interpretation is also qualitatively consistent with the role of large-scale fluctuations in the outer secondary peak proposed by \citet{Jimenez24}.

Further support for this possibility can be found from a physical-space perspective.
Using instantaneous three-dimensional $u$-structures, \citet{Yoon20} explicitly separated wall-attached and wall-detached motions according to their minimum distance from the wall.
They showed that tall wall-detached structures are geometrically self-similar and contribute predominantly to the detached streamwise intensity in the outer region.
In particular, the intensity carried by these structures exhibits a distinct outer maximum, which is further enhanced under an adverse pressure gradient (APG).
Although the wall-detached structures identified by \citet{Yoon20} cannot be directly equated with the residual defined here, these observations provide structural evidence that energetic motions without a direct near-wall connection can make a localized contribution to the outer-region intensity.
In this respect, such motions offer one possible physical interpretation of the weak outer-region hump observed in $\langle uu\rangle_r^+$.
The enhancement of these wall-detached self-similar motions under an adverse pressure gradient further suggests that APG flows provide a useful configuration for examining the structural origin of outer-peak formation.
Applying the present coherence-based decomposition to APG flows may help determine whether the enhanced outer peak is accompanied by an increased contribution from wall-incoherent motions, thereby providing a test of the mechanism suggested by the present residual analysis.

\section{Further discussion}
\label{sec:6}
\subsection{A possible geometrical interpretation of the most active location}
\label{sec:6.1}
A notable observation in the present work is that the most active location of an individual wall-coherent motion remains approximately fixed at $y_a^*=y_a/y_l\simeq0.38$--$0.40$ for both the streamwise and wall-normal velocity components over the $Re_\tau$ range considered (figure~\ref{Fig04}).
The origin of this value is not immediately evident.
Here, we provide a possible geometrical interpretation based on the hierarchical organisation of uniform momentum zones (UMZs), which appear in instantaneous flow fields as regions of relatively uniform streamwise momentum separated by thin shear layers \citep{Meinhart95}.

\citet{Perry82} introduced a hierarchy of geometrically similar eddies to sustain the logarithmic mean-velocity profile and approximately constant Reynolds shear stress over an increasingly broad wall-normal range as the Reynolds number increases. 
UMZs provide a useful physical link between the discrete and continuous descriptions of this hierarchy. 
In the framework of \citet{Perry82}, randomness and jitter smear the discrete hierarchy into a continuous distribution of scales, while the instantaneous zonal structure associated with UMZs provides a possible geometrical representation of the underlying hierarchy, as discussed in \citet{Hwang18}.
Interestingly, similar hierarchical structures also arise from theoretical analyses. \citet{Busse70} obtained a cascade of discrete scales from a variational analysis of momentum transport, while \citet{Klewicki14,Klewicki21} derived a hierarchy of scaling layers from an invariant form of the mean momentum equation.
These results suggest that a hierarchical organisation of wall turbulence may have a broader physical basis beyond its original formulation in the AEM.

As discussed in \S~\ref{sec:3}, the present wall-coherent motions share statistical characteristics with the instantaneous wall-attached $u$ structures identified by \citet{Hwang18}.
These structures are composed of multiple UMZs, whose number increases with their wall-normal extent.
This observation suggests that the internal organisation of a large wall-attached structure may be interpreted as a hierarchy of UMZs in the context of the AEM \citep{Marusic01,deSilva16}.
Within the mean-momentum-balance framework described above, these scaling layers are associated with UMZs in the inertial sublayer \citep{Priyadarshana07}, with their thicknesses forming a geometric progression \citep{Klewicki14,Bautista19}.
More recently, \citet{Kim26} showed that wall-scaled UMZs, whose thicknesses and velocity jumps scale with $y$ and $u_\tau$, respectively, exhibit a geometric progression consistent with the hierarchy proposed by \citet{Klewicki14,Klewicki21}.
Motivated by these observations, we idealise a wall-coherent motion of height $y_l$ as a hierarchy of UMZs whose thicknesses follow a geometric progression.

Let $W_i$ denote the thickness of the $i$th UMZ and $y_i$ its lower wall-normal edge, with $i$ increasing away from the wall.
Within the scaling-layer framework, these quantities satisfy
\par
\begin{linenomath}
\begin{equation}
y_{i+1}=y_i+W_i,
\qquad
y_i=\phi_c W_i,
\qquad
\frac{W_{i+1}}{W_i}
=
\frac{\phi_c+1}{\phi_c}.
\label{eq6.1}
\end{equation}
\end{linenomath}
\par\noindent
The mean-momentum-balance framework \citep{Klewicki14,Klewicki21} suggests that, in the asymptotic high-Reynolds-number limit, $\phi_c$ approaches the golden ratio, $\phi_c\rightarrow(1+\sqrt{5})/2\approx1.62$, which satisfies $\phi_c+1=\phi_c^2$.
Consistent with this relation, \citet{Kim26} found that the lower and upper edges of wall-scaled UMZs, denoted by $y_L$ and $y_U$, respectively, approximately satisfy $y_U/y_L\simeq1.62$.
For the golden-ratio value of $\phi_c$, the thickness ratio in \eqref{eq6.1} reduces to $W_{i+1}/W_i=\phi_c$, giving the geometric progression $W_i=W_1\phi_c^{\,i-1}$.

For a hierarchy consisting of $n$ UMZs following the geometric progression obtained above, the total thickness of the stacked UMZs is obtained by summing their individual thicknesses.
The locations $y_i$ correspond to the internal interfaces separating successive UMZs.
These interfaces (or vortical fissures) are dynamically important because significant momentum exchange occurs across them \citep{Adrian00,Priyadarshana07,deSilva17,Heisel20,Kim26}.
Since the most active location identified here corresponds to the region of strongest wall-normal velocity contribution, we seek a characteristic location representative of these internal interfaces.
As a simple geometrical measure, we weight each $y_i$ by the thickness $W_i$ of the corresponding UMZ.
Thus,
\par
\begin{linenomath}
\begin{subequations}
\label{eq6.2}
\begin{equation}
H_n
=
\sum_{i=1}^{n}W_i
=
W_1\frac{\phi_c^n-1}{\phi_c-1}
=
W_1(\phi_c^{n+1}-\phi_c),
\label{eq6.2a}
\end{equation}
\begin{equation}
h_n
=
\frac{\sum_{i=1}^{n}W_i y_i}{H_n}
=
\phi_c W_1
\frac{\phi_c^n+1}{\phi_c+1}.
\label{eq6.2b}
\end{equation}
\end{subequations}
\end{linenomath}
\par\noindent
Here, $H_n$ represents the total thickness occupied by the $n$ UMZs, whereas $h_n$ represents their thickness-weighted characteristic interface location.
Since the first UMZ begins at $y_1=\phi_c W_1$, the upper edge of the hierarchy is located at $y_{n+1}=y_1+H_n=W_1\phi_c^{n+1}$.
Under the present idealised representation, we associate this upper edge with the coherence height of the wall-coherent motion, such that $y_l\simeq y_{n+1}$.

Under this interpretation, we take $y_a\simeq h_n$.
The resulting normalized most active location is therefore
\par
\begin{linenomath}
\begin{equation}
y_a^*
=
\frac{y_a}{y_l}
\simeq
\frac{h_n}{y_{n+1}}
=
\frac{1}{\phi_c^2}
\left(1+\phi_c^{-n}\right)
\;\longrightarrow\;
\frac{1}{\phi_c^2}
\simeq0.382
\qquad (n\rightarrow\infty).
\label{eq6.3}
\end{equation}
\end{linenomath}
\par\noindent
The convergence in \eqref{eq6.3} is relatively rapid, giving $y_a^*\simeq0.403$ for $n=6$ and $y_a^*\simeq0.385$ for $n=10$, in agreement with the observed range $y_a^*\simeq0.38$--$0.40$ in figure~\ref{Fig04}.
Figures~\ref{Fig04}($e$,$f$) also suggest a weak Reynolds-number dependence, with $y_a^*$ decreasing from approximately $0.40$ at $Re_\tau=930$ towards $0.38$--$0.39$ at $Re_\tau=3000$ and $6000$.
Since the number of UMZs was observed to increase with Reynolds number \citep{deSilva16,Kim26}, this trend is qualitatively consistent with an increase in the number of layers $n$ in the present idealised hierarchy.
Equation~\eqref{eq6.3} then predicts a progressive approach towards $\phi_c^{-2}$ with increasing Reynolds number.

It is also worth noting that \citet{Klewicki21} related the von K\'arm\'an constant to the same hierarchy parameter through $\kappa=\phi_c^{-2}$, which approaches $\kappa\simeq0.382$ as $\phi_c$ tends to the golden ratio.
Thus, within the present idealised interpretation, the normalized most active location and the von K\'arm\'an constant share the same asymptotic limit, $y_a^*\rightarrow\phi_c^{-2}=\kappa$.
This correspondence suggests a possible connection between the wall-normal organisation of an individual wall-coherent motion and the hierarchical structure underlying the logarithmic mean velocity.

It should be emphasised, however, that this argument provides a geometrical interpretation rather than a dynamical derivation of $y_a^*$.
In particular, the association of $y_a$ with the thickness-weighted characteristic location of the hierarchical interfaces, $y_a\simeq h_n$, remains an assumption.
Moreover, the wall-coherent motions extracted in the present analysis have not been explicitly decomposed into individual UMZs.
The construction also considers only the relative geometrical arrangement of the hierarchy and does not explicitly account for its Reynolds-number-dependent inner and outer bounds.
Nevertheless, the close agreement between the asymptotic value $\phi_c^{-2}\simeq0.382$ and the observed $y_a^*\simeq0.38$--$0.40$ suggests that the hierarchical organisation of UMZ interfaces may provide a plausible structural interpretation of the most active location of a wall-coherent motion.

The numerical proximity of $y_a^*$ to the von K\'arm\'an constant also suggests a possible connection with the statistical AE formulation of \citet{Woodcock15}.
Using Campbell's theorem, they derived the von K\'arm\'an constant analytically in terms of the first-order streamwise contribution of a representative AE.
Their expression includes a finite-Reynolds-number correction that rapidly approaches unity, leading to only a weak Reynolds-number dependence of $\kappa$ at sufficiently high $Re_\tau$.
This result suggests a possible test of the present interpretation.
If the first-order contribution associated with the wall-coherent motions identified here can be extracted alongside the second-order intensity functions considered in this study, the analytical relation proposed by \citet{Woodcock15} could be examined directly within the present single-eddy framework.

The geometrical argument above is essentially algebraic and does not explain the dynamical mechanism that selects the most active location.
If the most active location is interpreted as the region where an individual motion most efficiently extracts energy from the mean flow, its origin should ultimately be understood from a dynamical rather than purely geometrical perspective.
In this respect, critical-layer dynamics may provide a useful perspective, since the interaction responsible for sustaining a coherent motion becomes concentrated near the critical layer where its phase speed matches the local streamwise velocity. 
Recently, \citet{Song26} showed that multiscale quasi-time-periodic coherent states can support a hierarchy of critical layers and associated vortical structures whose scales decrease towards the wall. 
Whether the present value $y_a^*\simeq \kappa$ can be related to such a critical-layer mechanism remains an important question for future study.

\subsection{Logarithmic variations and the Townsend--Perry constant}
\label{sec:6.2}
As shown in \S\S~\ref{sec:3.4} and \ref{sec:3.5}, the present single-eddy decomposition reveals two distinct logarithmic tendencies in the cumulative contributions to the streamwise turbulence intensity.
The first arises from the accumulation of the inactive footprints of wall-coherent motions, represented by $\langle uu\rangle_{ia}^+$, and is characterized by a logarithmic coefficient of approximately $A_i\simeq0.65$ over $30\lesssim y^+\lesssim0.025R^+$.
The second appears farther from the wall in the broader cumulative contribution $\langle uu\rangle_c^+$, for which the logarithmic indicator reaches approximately $\Xi_c\simeq1.3$ over $0.14\lesssim y/R\lesssim0.25$.
Although both behaviours originate from the inverse-height scaling of the single-eddy intensity, their physical origins are different: the former reflects the accumulation of inactive footprints, whereas the latter results from integrating the active portion of the hierarchy up to the finite outer cutoff at $y_l\simeq0.5R$.
These observations motivate a closer examination of how the logarithmic coefficients identified here are related to the Townsend--Perry constant $A_1$ of the streamwise turbulence intensity in \eqref{eq1.3}.

The coefficient $A_i\simeq0.65$ associated with the inactive contribution is considerably smaller than the value $A_1\simeq1.26$ obtained from the total streamwise turbulence intensity at high Reynolds numbers \citep{Marusic13,Samie18}.
The total intensity, however, contains contributions from motions spanning a broad range of scales and therefore does not specifically represent the inactive footprints of the self-similar wall-attached hierarchy \citep{Yoon20,Hwang20,Baars20}.
Using spectral filtering to isolate the component associated predominantly with self-similar wall-attached motions, \citet{Baars20b} obtained a smaller value of $A_1\simeq0.98$, although, as noted by the authors, the extracted component cannot be regarded as a pure representation of self-similar AEs.
A similar value was obtained by \citet{Deshpande21} after further isolating the contribution attributed to inactive self-similar AEs.
As discussed in \S~\ref{sec:2.3}, however, their cumulative decomposition may still include contributions from wall-coherent motions with $y_l=O(y)$ that remain active at the observation location.
The smaller value of $A_i$ obtained here may therefore reflect, at least in part, the more restrictive isolation of the inactive footprints of the wall-coherent hierarchy.

The logarithmic region reported in previous studies, however, extends farther from the wall than that found here for $\langle uu\rangle_{ia}^+$, with its outer limit typically reaching $y/R\simeq0.15$ \citep{Marusic13,Samie18}.
According to the present single-eddy results, the inactive region is confined approximately to $y^*=y/y_l\lesssim0.1$.
For an inactive footprint to contribute at $y/R=0.15$, the corresponding wall-coherent motion would require $y_l/R\gtrsim1.5$, which exceeds the maximum possible wall-normal extent in the pipe.
The logarithmic variation observed at these outer-scaled wall distances therefore cannot be explained solely by the accumulation of the deep inactive footprints identified in the present analysis.

By contrast, the broader cumulative contribution $\langle uu\rangle_c^+$ exhibits a logarithmic coefficient of $\Xi_c\simeq1.3$, much closer to the Townsend--Perry constant $A_1\simeq1.26$ reported for the total streamwise turbulence intensity.
As discussed in \S~\ref{sec:3.5}, this logarithmic variation results from integrating the active portion of the wall-coherent hierarchy up to its finite outer cutoff at $y_l\simeq0.5R$.
Equation~\eqref{eq3.10} further shows that an approximately constant logarithmic slope can arise when the single-eddy prefactor $A_a$ varies only weakly over the range sampled by the moving lower bound $y^*=2y/R$.
Hence, these results suggest that the logarithmic variation observed in the total intensity at finite outer-scaled wall distances may be influenced by the finite outer cutoff of the active hierarchy, rather than arising solely from the accumulation of inactive footprints.

The present interpretation also provides a possible connection to the logarithmic variation previously observed in wall-attached structures extracted directly from instantaneous flow fields \citep{Hwang18,Hwang19,Hwang20}. 
In those studies, the streamwise velocity fluctuations $u$ were retained only within the identified structures, such that the contribution of each structure was naturally truncated at its boundary.
Importantly, $u$ changes rapidly near the boundaries of the identified structures \citep{Hwang22}, making this geometrical truncation a natural approximation to the spatial cutoff of their contribution.
Since taller wall-attached structures were found to contain the collective contributions of smaller hierarchical motions, this geometrical truncation may be reinterpreted as an effective cutoff of the active portion of the hierarchy as the observation location moves away from the wall. 
In this respect, the resulting logarithmic variation can be viewed as a physical-space manifestation of the progressive cutoff of active contributions within the hierarchy identified in the present analysis.

The finite-$y/R$ effect associated with the active hierarchy can also be related to the spectral-scaling interpretation of \citet{Hwangy22}.
Although their analysis considered this correction primarily in the context of the inactive contribution of AEs, both the active and inactive portions identified here exhibit the underlying inverse-height scaling, $I_{uu}^+\sim(y_l^+)^{-1}$, at fixed $y^*$ over their respective coherence-height ranges.
The residual $y/R$ dependence of the spectrum discussed by \citet{Hwangy22} may therefore reflect how different portions of the AE hierarchy are sampled at finite $y/R$.
In Part~2, this connection will be examined further through the spectral contributions of individual wall-coherent motions.

It should be emphasised, however, that the above discussion concerns only wall-coherent motions.
Wall-incoherent motions also possess locally active portions and make substantial contributions to the streamwise turbulence intensity in the outer region.
Since these motions are not resolved according to their wall-normal coherence heights in the present analysis, their role in the logarithmic variation remains to be examined.


\section{Concluding remarks}
\label{sec:7}
In the present study, we have investigated the contribution of individual wall-coherent motions to the streamwise turbulence intensity in turbulent pipe flow at $Re_\tau=930$, $3000$ and $6000$.
Using spectral linear stochastic estimation together with differencing between cumulative wall-coherent contributions, we isolated motions associated with a prescribed wall-normal coherence height $y_l$ and obtained their single-eddy intensity functions.
To distinguish the active and inactive portions in accordance with Townsend's original description, we compared the streamwise and wall-normal single-eddy intensity functions, $I_{uu}(y;y_l)$ and $I_{vv}(y;y_l)$.
The main findings are summarised as follows.

\begin{enumerate}[
    label=(\roman*),
    labelsep=0.5em,
    leftmargin=2.5em,
    itemsep=0.5\baselineskip,
    parsep=0pt,
    topsep=0.5\baselineskip,
    partopsep=0pt
]

\item
The wall-normal behaviour of an individual wall-coherent motion was characterised using $I_{uu}$ and $I_{vv}$.
When expressed in terms of the eddy-scaled wall-normal location ($y^*=y/y_l$), the profiles of both components exhibit reasonable collapse, with their strongest contributions occurring at $y_a^*\simeq0.38$--$0.40$.
Towards the wall, $I_{vv}$ decreases rapidly while $I_{uu}$ remains finite, revealing a near-wall inactive portion.
The inverse-height scaling, $I_{uu}^+\sim(y_l^+)^{-1}$, consistent with the classical AE hierarchy \citep{Townsend76,Perry82,Perry86}, persists from the active portion into the inactive portion, although over different coherence-height ranges.
These observations were used to define the active and inactive regions summarised in \eqref{eq3.3a} and \eqref{eq3.3b}, respectively.
Interestingly, the observed $y_a^*$ is close to the von K\'arm\'an constant.
As discussed in \S~\ref{sec:6.1}, this value may have a geometrical interpretation based on a hierarchy of UMZ thicknesses following a geometric progression proposed by \citet{Klewicki14}.

\item
The cumulative active and inactive contributions were examined by integrating the single-eddy intensity over the corresponding regions, with the inverse-height scaling, $I_{uu}^+\sim(y_l^+)^{-1}$, implying approximately equal contributions from equal logarithmic intervals of $y_l$.
For the active portion, both bounds scale with $y$, yielding an approximately constant contribution across the inertial sublayer, whereas the accumulation of inactive footprints produces a logarithmic variation with $A_i\simeq0.65$, consistent with the classical AEH.
This inactive logarithmic variation is confined to $y/R\lesssim0.025$, because the inactive portion occurs at $y^*\lesssim0.1$ and its inverse-height scaling extends only to $y_l\lesssim0.25R$.
A broader integration reveals a second logarithmic tendency over $0.14\lesssim y/R\lesssim0.25$, arising entirely from integrating the active hierarchy up to its finite outer cutoff at $y_l\simeq0.5R$.
The corresponding logarithmic indicator is $\Xi_c\simeq1.3$, close to the Townsend--Perry constant $A_1\simeq1.26$, as discussed further in \S~\ref{sec:6.2}.
Thus, the two logarithmic tendencies have distinct physical origins despite arising from the same underlying single-eddy scaling and bear a qualitative resemblance to the two invariant regions of the $u$ p.d.f. recently reported by \citet{Tsuji26} and \citet{Tsuji26tsfp}.

\item
At the near-wall peak location, $y^+=15$, the single-eddy intensity exhibits a steeper coherence-height dependence, $I_{uu}^+\sim(y_l^+)^{-(1+\beta)}$ with $\beta\simeq1/4$, reflecting the additional attenuation of the near-wall inactive footprint as the fixed observation location samples progressively smaller $y^*$ with increasing $y_l$.
The near-wall peak intensity was decomposed into residual small-scale, predominantly active, inactive-footprint and outer-scale contributions.
The first two are expected to approach $Re_\tau$-independent limits, whereas the inactive-footprint contribution approaches a finite asymptote with a $Re_\tau^{-\beta}$ defect and the outer-scale contribution is governed by the outer-cutoff scaling, leading to $\langle uu\rangle_{\max}^+=A-BRe_\tau^{-1/4}$.
The exponent $\beta$ was interpreted using the equilibrium-layer estimate $\varepsilon_l\sim u_\tau^3/y_l$, which gives $\eta_l^+\sim(y_l^+)^{1/4}$.
Assuming that the attenuation scales as $\delta_\nu/\eta_l$ then yields $\beta=1/4$, providing a phenomenological explanation for the observed exponent.
This interpretation is consistent with Bradshaw's description of turbulent-energy diffusion and dissipation associated with inactive motions \citep{Bradshaw67}.
Thus, although the resulting $Re_\tau^{-1/4}$ defect coincides with that proposed by \citet{Chen21}, the present interpretation is based on the eddy-height-dependent dissipative scale of the AE hierarchy rather than on an assumed limiting wall dissipation and outer-scale transport of near-wall turbulence.

\item
The decomposition also provides insight into the development of the outer-region streamwise intensity.
The contribution from wall-coherent motions with $y_l^+\geq200$ increases with $Re_\tau$ and extends progressively farther from the wall, thereby promoting the formation of an outer shoulder; however, its maximum remains near $y^+\simeq100$ and is substantially influenced by the lower coherence-height cutoff, rather than representing a distinct outer secondary peak.
In contrast, the residual contribution exhibits a weak localized outer-region feature that becomes more pronounced from $Re_\tau=3000$ to $6000$.
If this tendency persists at higher $Re_\tau$, a distinct outer secondary maximum may emerge through the superposition of this localized residual contribution and the more broadly distributed wall-coherent contribution.
Since the residual also contains wall-incoherent motions, including potentially large-scale motions with locally active portions, such motions may play an important role in the eventual formation of the outer secondary peak.
\end{enumerate}

Taken together, the present results provide a scale-resolved link between the wall-normal behaviour of individual wall-coherent motions and several statistical features associated with the AEH.
A broader range of Reynolds numbers is required to establish more firmly the scaling ranges and their bounds summarised in \eqref{eq3.3}.
In particular, the lower bound associated with viscous effects may be expected to scale as $Re_\tau^{1/2}$, but the present Reynolds-number range is too limited to verify this dependence conclusively.
Further work should also extend the present framework to wall-incoherent motions and the spanwise velocity component.
In particular, examining the convection velocity of individual wall-coherent motions may provide further dynamical insight into their active and inactive portions.
In Part~2, the one- and two-dimensional spectral contributions of individual wall-coherent motions will be examined to investigate further the spectral origin of the single-eddy scaling and its finite-$y/R$ dependence.\\

\noindent \textbf{Acknowledgements}
\noindent This work was supported by the National Research Foundation of Korea (NRF) grant funded by the Korea government (MSIT) (Grant No. RS-2023-00211896) and was partially supported by the Supercomputing Centre (KISTI).\\

\noindent \textbf{Declaration of interests}
\noindent The author reports no conflict of interest.

\appendix

\section{Derivation of the near-wall defect scaling}
\label{app:peak_scaling}

The Reynolds-number dependence of the near-wall streamwise intensity can be obtained by integrating the single-eddy model introduced in \eqref{eq4.3}.
At $y^+=15$, 
\par
\begin{linenomath}
\begin{equation}
I_{uu}^+(15;y_l^+)
\simeq
K_{15}(y_l^+)^{-(1+\beta)}
G(y_l/R),
\label{eqA1}
\end{equation}
\end{linenomath}
\par\noindent
where $K_{15}$ is the prefactor, $\beta$ represents the additional attenuation of the near-wall inactive footprint, and $G$ accounts for the outer-scale cutoff.
For the present data, $\beta\simeq1/4$, while $G\simeq1$ over $y_i^+\lesssim y_l^+\lesssim c_oR^+$, with $y_i^+=200$ and $c_o\simeq0.25$.

The near-wall inactive-footprint contribution is therefore
\par
\begin{linenomath}
\begin{equation}
\begin{aligned}
T_i^+
&=
\int_{y_i^+}^{c_oR^+}
K_{15}(y_l^+)^{-(1+\beta)}
\,\mathrm{d}y_l^+
\\
&=
\frac{K_{15}}{\beta}
\left[
(y_i^+)^{-\beta}
-
(c_oRe_\tau)^{-\beta}
\right]
\\
&\equiv
T_{i,\infty}^+
-
D_iRe_\tau^{-\beta},
\end{aligned}
\label{eqA2}
\end{equation}
\end{linenomath}
\par\noindent
where
$T_{i,\infty}^+=(K_{15}/\beta)(y_i^+)^{-\beta}$ and $D_i=(K_{15}/\beta)c_o^{-\beta}$.
Thus, $T_i^+$ increases with $Re_\tau$ but approaches a finite asymptotic value ($T_{i,\infty}^+$) with a defect proportional to $Re_\tau^{-\beta}$.

For $y_l/R\gtrsim c_o$, the outer-cutoff function must be retained.
Introducing the outer-scaled coherence height $Y=y_l/R=y_l^+/Re_\tau$, the outer-scale contribution becomes
\par
\begin{linenomath}
\begin{equation}
\begin{aligned}
T_o^+
&=
\int_{c_oR^+}^{R^+}
K_{15}(y_l^+)^{-(1+\beta)}
G(y_l/R)
\,\mathrm{d}y_l^+
\\
&=
K_{15}Re_\tau^{-\beta}
\int_{c_o}^{1}
Y^{-(1+\beta)}G(Y)
\,\mathrm{d}Y
\\
&\equiv
D_oRe_\tau^{-\beta},
\end{aligned}
\label{eqA3}
\end{equation}
\end{linenomath}
\par\noindent
where
$
D_o
=
K_{15}
\int_{c_o}^{1}
Y^{-(1+\beta)}G(Y)
\,\mathrm{d}Y.
$
Provided that $K_{15}$ and the outer-scaled form of $G(Y)$ approach Reynolds-number-independent limits, $D_o$ becomes asymptotically Reynolds-number independent.

Combining these results with the residual and lower-coherence-height contributions gives
\par
\begin{linenomath}
\begin{equation}
\begin{aligned}
\langle uu\rangle_{\max}^+
&=
T_s^+ + T_a^+ + T_i^+ + T_o^+
\\
&=
T_s^+ + T_a^+ + T_{i,\infty}^+
-
(D_i-D_o)Re_\tau^{-\beta}.
\end{aligned}
\label{eqA4}
\end{equation}
\end{linenomath}
\par\noindent
As discussed in \S~\ref{sec:4}, $T_s^++T_a^+$ is expected to approach a Reynolds-number-independent limit because it is dominated by small-scale and predominantly active contributions.
If $K_{15}$ and $G$ also approach Reynolds-number-independent limits, \eqref{eqA4} reduces to the defect form in \eqref{eq4.6}, with $A=(T_s^++T_a^+)+T_{i,\infty}^+$ and $B=D_i-D_o$.
The possible finite-Reynolds-number dependence of $K_{15}$ is discussed in \S~\ref{sec:4.2}.
For the observed $\beta\simeq1/4$, the leading defect is proportional to $Re_\tau^{-1/4}$.

\bibliographystyle{jfm}
\bibliography{bbl_data}

\end{document}